\documentclass{aa}  
\usepackage{graphicx}
\usepackage{txfonts}
\usepackage{hyperref}
\hypersetup{
    colorlinks=true,
    linkcolor=blue,
    filecolor=magenta,      
    urlcolor=blue,
    citecolor=blue
    }
\usepackage{amsmath}
\usepackage[version=4]{mhchem}
\usepackage{gensymb}
\usepackage{chemfig}
\usepackage{comment}

\usepackage{natbib}
\bibpunct{(}{)}{;}{a}{}{,}

\begin{document}

   \title{Three dimensional temporal evolution of photochemical hazes  \\ in exoplanet atmospheres}

   \subtitle{I. Description and test application to HD 189733b}

   \author{Elspeth K.H. Lee\inst{1}\thanks{corresponding author: elspeth.lee@unibe.ch}, Maria E. Steinrueck\inst{2}, Kazumasa Ohno\inst{3}, Diana Powell\inst{2} and Xi Zhang\inst{4}
        }

   \institute{$^{1}$Center for Space and Habitability, University of Bern, Gesellschaftsstrasse 6, 3012 Bern, Switzerland\\
   $^{2}$Department of Astronomy \& Astrophysics, University of Chicago, Chicago, IL 60637, USA\\
   $^{3}$Division of Science, National Astronomical Observatory of Japan, 2-21-1 Osawa, Mitaka-shi, Tokyo, Japan \\
   $^{4}$Department of Earth and Planetary Sciences, University of California, Santa Cruz, CA 95064, USA}

  \date{Received 27 April 2026 / Accepted 17 June 2026}
 
  \abstract
   {The formation and global spatial distribution of photochemically produced haze particles remain a key process in exoplanet atmospheres for understanding their observed properties.}
   {We aim to develop a flexible haze particle formation and evolution model suitable for time-dependent exoplanet atmosphere simulations.} 
   {Inspired by recent 2D photochemical modelling efforts, we include a simple activation timescale mechanism to emulate a delayed formation of solid haze particles.
    We couple our new microphysical haze formation scheme, {\tt mini-haze}, to the {\tt Exo-FMS} general circulation model (GCM) and simulate an idealised HD 189733b case study to examine the 3D spatial distribution and sizes of haze particles.}
   {Our results suggest that for our chosen haze formation efficiency, particles do not grow beyond $\sim$ 30\,nm, in line with previous detailed 1D modelling.
    We find the haze spatial distribution follows the vertical velocity structure of the atmosphere, with equatorial convergence patterns of material deeper in the atmosphere at $\sim$ 10$^{-2}$\,bar.
    The resulting global distribution leads to enhanced haze opacity at the east and west limbs of the atmosphere. 
    In our test cases, radiative feedback from haze opacity can strongly affect the temperature-pressure structures in the upper atmosphere depending on the production rate.
    Our synthetic spectra results suggest that longer haze-production timescales give rise to stronger haze opacity effects on the observed transmission spectra compared to short-timescale dayside formation, but the stronger thermal feedback from nightside formation leads to an overall larger dayside emission flux.}
   {Our current simulations represent a step towards investigating self-consistent haze formation and evolution with chemical feedback effects in 3D, and can be readily applied to other objects of interest, such as sub-Neptune atmospheres.}


   \titlerunning{{\tt mini-haze}}
   \authorrunning{Lee et al.}

   \maketitle

\nolinenumbers

\section{Introduction}

The presence of photochemically produced aerosol particles is ubiquitous in Solar System objects with an appreciable atmosphere.
For exoplanet science, understanding the formation and spatial distribution of photochemically produced haze  particles remains a key challenge for characterisation of the physical and chemical processes that occur in these objects.
In exoplanet literature, generally, `aerosol' refers to a generic atmospheric particle, `cloud' a particle that derives from condensation of material, and `haze' a particle that originates from photochemical processes.

Increasing evidence from observations of sub-Neptune-sized planets suggests that many exhibit strong cloudiness and haziness characteristics.
\citet{Kreidberg2014} discovered that the canonical sub-Neptune GJ 1214b had a very flat transmission spectrum in the HST WFC3 near-infrared wavelength range, indicative of a thick high altitude cloud or photochemical haze component at the transmission limbs.
Recently, GJ 1214b was the target of several JWST observational campaigns.
\citet{Kempton2023} used MIRI LRS to observe a thermal phase curve of GJ 1214b. 
They found a highly reflective atmosphere, with Bond albedo $A_{\rm B}$ $\sim$ 0.51, and low thermal flux on the nightside, suggesting a thick, global haze or cloud component.
\citet{Malsky2025} reanalysed the GJ 1214b MIRI phase curve data and performed general circulation model (GCM) simulations including a haze and cloud component. 
They found that a slightly lower Bond albedo of 0.42 best fits their phase-curve data.
\citet{Schlawin2024} used NIRSpec G395H to measure two transits of GJ 1214b.
They found tentative evidence of \ce{CH4} and \ce{CO2} absorption, able to be detectable above the strong aerosol component.
Both these studies suggested a highly super-solar metallicity atmosphere ($\sim$ 100--3000 $\times$ solar), with a high molecular weight atmosphere.
Recent JWST observations of sub-Neptunes that have evidence of haze particles in their atmospheres include TOI-836c \citep{Wallack2024}, GJ 3090b \citep{Ahrer2025}, and TOI-776c \citep{Teske2025}.
\citet{Brande2024} collated the available HST WFC3 transmission spectrum data for exo-Neptunes.
They suggested a continuum of haze and cloudy atmospheres between $T_{\rm eq}$ = 200\,K and 1000\,K, with clear models preferred at lower $<$ 500\,K and higher $>$ 700\,K equilibrium temperature objects, and an aerosol component in the atmosphere between this range.

Significant effort has gone into simulating the effects of photochemically produced haze in 1D vertical column modelling.
\citet{Morley2013} and \citet{Morley2015} examined the haze structures of GJ 1214b and sub-Neptunes respectively.
They found that models that convert around 10\% of the precursor species mass into haze particles can qualitatively reproduce the flat transmission spectra of GJ 1214b seen by \citet{Kreidberg2014}.
\citet{Lavvas2017} applied their kinetic chemistry and haze formation scheme to the hot Jupiter HD 189733b.
They found that small, nanometre-sized haze particles could explain the super-Rayleigh slope seen in the HST transmission spectrum \citep[e.g.][]{Sing2016}.
\citet{Kawashima2018} coupled a photochemical kinetics model to a haze monomer production scheme and collisional growth bin model and investigated the effect of the size distributions on the transmission spectra of warm-Neptunes.
\citet{Gao2020} suggested a regime shift between cloud-dominated and haze-dominated opacity features in transmission spectra occurring at approximately $T_{\rm eff}$ $\approx$ 950\,K.
\citet{OhnoKawashima2020} systematically investigated the impacts of hazes on optical spectral slopes and found that hazes could produce super-Rayleigh slopes under strong eddy diffusion.
\citet{Arfaux2024} combined haze and condensate microphysical processes to model the different transmission spectra produced on the limbs of the hot-Saturn WASP-39b.
\citet{Ohno2024} investigated the possibility of the formation of hazes composed of diamond in hot exoplanetary atmospheres by utilizing the theory of carbon vapour deposition established in the industry community.
\citet{OwenMurray25} studied the influence of radiative pressure on the dynamics of haze particles using a 2D equatorial band model.

Due to the newly released JWST data on GJ 1214b \citep{Kempton2023, Schlawin2024}, recent modelling studies have focused on exploring the haze properties of this sub-Neptune.
\citet{Gao2023} investigated super-solar and steam atmospheric environments scenario for GJ 1214b, using the {\tt CARMA} microphysical model to produce haze size distribution profiles for each scenario.
\citet{Lavvas2024} used a coupled chemistry-haze formation scheme, and found that efficient haze formation on GJ 1214b can best explain the transmission and emission properties of the planet.
\citet{Ohno2025} performed a suite of forward models across a wide range of metallicities, and found a metal enhanced atmosphere with photochemically produced haze can well fit the JWST measurements.

Haze particles have also been produced in laboratory settings that emulate hydrogen-rich exoplanet atmospheres.
Examples include \citet{Horst2018} and \citet{He2018} who measured haze production rates in conditions suitable for sub-Neptune atmospheric environments with various initial gas mixtures.
\citet{Fleury2019} observed the formation of hazes in high-temperature photochemistry experiments in the case of an atmosphere with a high C/O ratio. 
\citet{Moran2020} measured the complex composition of these laboratory-produced hazes using mass spectrometry, finding a multitude of nitrogenated-hydrocarbon components.
\citet{Yu2021} measured and calculated the surface tension of these laboratory made hazes, examining the potential for condensates to form on their surfaces.
\citet{Corrales2023} and \citet{He2024} provided measurements of optical properties of laboratory hazes produced under conditions relevant to exoplanets, and \citet{Huseby2025} and \citet{Pesciotta2026} studied how exposure to UV radiation and water changes these optical properties.

Despite progress in modelling hazes in 1D, only a few studies have investigated the global spatial distribution of haze particles in 3D using general circulation model (GCM) simulations.
For Solar System objects, GCM simulations including a haze component have been performed for Titan \citep{Lebonnois2012,Larson2015}, Pluto \citep{Bertrand2017}, and Archean Earth \citep{Mak2023}.
For Earth studies, sophisticated chemical-aerosol interaction schemes have been developed for GCMs, for example the {\tt GLOMAP} model \citep[e.g.][]{Mann2010}.
For exoplanet atmospheres in 3D, \citet{Cohen2024} applied a simple haze formation model to a GCM to simulate haze distributions on super-Earth planets.
\citet{Mak2024} simulated haze particle distributions on TRAPPIST-1e using GCM simulations.
Hot Jupiter GCM haze studies by \citet{Steinrueck2021} and \citet{Steinrueck2023} suggested the size properties of the haze particles drastically affects their spatial 3D distribution and radiative-feedback effects, which alter the haze opacity features in transmission spectra in a complex manner.
\citet{Mak2025} further studied the impacts of different haze compositions on the atmospheric dynamics for a broader sample of hot Jupiters.
\citet{Steinrueck2025} also simulated atmospheric dynamics on hazy sub-Neptunes to examine the possible degeneracy between hazes and atmospheric metallicity when interpreting thermal phase curves.

HD 189733A is an active K2V dwarf star. 
\citet{Bouchy_2005} discovered it hosts a hot Jupiter, HD 189733b, with radius $R_{\rm p}$ $\approx$ 1.12\,$R_{\rm Jup}$ and mass $M_{\rm p}$ $\approx$ 1.17\,$M_{\rm Jup}$ \citep{Addison2019}.
Its favourable transmission spectroscopy metric \citep[TSM;][]{Kempton_2018} has made it one of the most studied canonical hot Jupiters, with extensive transmission, emission and phase-curve coverage from HST STIS and WFC3 \citep[e.g.][]{Pont2013, Sing2016} and \textit{Spitzer} \citep[e.g.][]{Grillmair_2007, Agol_2010, Desert_2011, Knutson_2012}.
The active nature of the host star has, however, made characterisation of the optical scattering slope, and hence the aerosol component of the atmosphere, a challenge \citep[e.g.][]{Pont2013}.
In the JWST era, HD 189733b continues to be characterised in detail in both transmission \citep[e.g.][]{Fu2024} and dayside emission \citep[e.g.][]{Inglis2024}, motivating its use as the test case for our 3D haze modelling.

One limitation of previous studies is that they apply fixed haze properties, such as particle sizes, in GCM simulations, although atmospheric dynamics and haze evolution properties would interact with each other in reality.
With interest in understanding the formation and distribution of haze particles in exoplanet atmospheres, boosted by the release of recent JWST data on sub-Neptunes, a 3D holistic microphysical approach to the evolution of haze particles is highly warranted in order to understand the effects of haze on their atmospheric properties and observables.
To further invest in this important topic, we develop `{\tt mini-haze}', a generalised, microphysical time-dependent haze production, coagulation, and coalescence scheme for use in 1D-3D exoplanet atmosphere simulations.
In this first test case study, we couple our new module to an idealised canonical hot Jupiter HD 189733b case and investigate the 3D distribution of haze particles in the atmosphere and their impact on observational properties.
In Section \ref{sec:mini-haze}, we present details on the haze particle two-moment scheme.
In Section \ref{sec:prod}, we detail our parameterised haze production scheme, including our proposed haze formation timescale scheme.
In Sections \ref{sec:GCM} and \ref{sec:res}, we couple the new module to the {\tt Exo-FMS} GCM and simulate two canonical HD 189733b hot Jupiter test cases with different assumptions on the haze formation timescales.
In Section \ref{sec:pp}, we post-process our GCM results, producing synthetic transmission and dayside emission spectra.
Section \ref{sec:disc} contains a discussion of our results.
Section \ref{sec:conc} contains the conclusion of our study.

\section{The {\tt mini-haze} model}
\label{sec:mini-haze}

In this Section, we describe the `{\tt mini-haze}' photochemical haze evolution model for exoplanet atmosphere simulations.
Like other models in the `{\tt mini-}' series, {\tt mini-chem} \citep{Tsai2022} and {\tt mini-cloud} \citep{Lee2023b}, the philosophy behind {\tt mini-haze} is to enable time-dependent, fully coupled microphysical processes to be used in large-scale 3D hydrodynamic simulations, such as GCMs, while retaining computational feasibility. 
{\tt mini-haze}, and other {\tt mini-} codes, aim to be intermediate complexity models, trading off the computational expense of more complete models, for example, large chemical kinetic networks \citep{Tsai_2021}, while retaining good accuracy.
This enables GCMs to be performed for longer timescales, allowing multi-scale timescale feedback mechanisms to be assessed and explored.
{\tt mini-haze} is written in Fortran 90, following standard coding conventions, with {\tt DLSODE} \citep{Radhakrishnan1993} as the stiff ordinary differential equation (ODE) solver.
The source code for {\tt mini-haze} is publicly available online on GitHub\footnote{\url{https://github.com/ELeeAstro/mini_haze}}.

\subsection{Two-moment haze evolution method}

The moment method, also known as the bulk method, evolves integrated moments of the size distribution.
Typically, this only involves one to four moments \citep[e.g.][]{Ohno2017}, or more if different phases or mixed particle compositions are simulated \citep[e.g.][]{Helling2008}.
This is in contrast to bin or spectral models which directly calculate the fluxes into and out of pre-defined particle size or mass bins, which can number up to $\sim$ 40 bins or more \citep[e.g.][]{Gao2018, Adams2019, Lavvas2024}.
This makes the moment method computationally efficient and generally more suitable for coupling to large scale 3D atmospheric models, such as GCMs, with current computational resources.
The moments, $M^{(k)}$ [g$^{k}$ cm$^{-3}$],  of the particle mass distribution are given by
\begin{equation}
    M^{(k)} = \int_{0}^{\infty}m^{k}f(m)dm,
\end{equation}
where $k$ is the integer or non-integer moment power, $m$ [g] the mass of the particle and $f(m)$ [cm$^{-3}$ g$^{-1}$] the particle mass distribution. 

In this study, we use the diagnostic variables, total number density $n_{\rm h}$ [cm$^{-3}$] ($k$ = 0; zeroth moment) of the haze particles and the mass density $\rho_{\rm h}$ [g cm$^{-3}$] ($k$ = 1; first moment) as the two moments, following closely the approach of \citet{Ohno2018}, \citet{Ohno2020} and \citet{Lee2025a}. 
We assume compact spheres for the haze particles, avoiding modifications to the rate equations from fluffy aggregate or fractal geometries \citep[e.g.][]{Ohno2020}.
We do not include any interaction between haze particles and potential condensate cloud materials in the current model.

The local time evolution equations for the moments are given as
\begin{equation}
    \frac{dn_{\rm h}}{dt} = \dot{P}_{\rm h}\left(\frac{\rho_{\rm a}}{V_{0}\rho_{\rm d}}\right) + \dot{L}_{\rm coll} + \dot{L}_{\rm deco},
\end{equation}
and
\begin{equation}
  \frac{d\rho_{\rm h}}{dt} = \dot{P}_{\rm h}\rho_{\rm a} + m_{\rm h}\dot{L}_{\rm deco},
\end{equation}
where $\dot{P}_{\rm h}$ [g g$^{-1}$ s$^{-1}$ $\equiv$ s$^{-1}$] is the mass mixing ratio production rate of haze particle monomers, $\rho_{\rm a}$ [g cm$^{-3}$] the atmospheric mass density, $V_{0}$ [cm$^{3}$] the volume of the haze particle monomers derived from a monomer radius $r_{0}$ [cm],  $\rho_{\rm d}$ [g cm$^{-3}$] the bulk density of the haze particle, $\dot{L}_{\rm coll}$ [cm$^{-3}$ s$^{-1}$] the loss of particle number density due to particle-particle collisions and $\dot{L}_{\rm deco}$ [cm$^{-3}$ s$^{-1}$], the loss due to thermal decomposition of haze particles.
Total mass is conserved during collisions, and so the first moment does not contain a collisional loss term.
Our formulation assumes `hit-and-stick' collisions, ignoring other effects such as bouncing or collisions that can fragment haze particles.
In addition, we do not consider charged surface effects that can alter the effective collision rates \citep[e.g.][]{Lavvas2010}.

Representative mean values of the mass distribution can be derived from the moment values. 
The number-weighted mean mass of the haze particles, $m_{\rm h}$ [g], is given by
\begin{equation}
\label{eq:mh}
    m_{\rm h} = \frac{\rho_{\rm h}}{n_{\rm h}},
\end{equation}
which is the ratio between the first and zeroth moment.
The representative radius for this number-weighted mass, $r_{\rm h}$ [cm], is
\begin{equation}
\label{eq:rv}
    r_{\rm h} = \left(\frac{3m_{\rm h}}{4\pi\rho_{\rm d}}\right)^{1/3}
\end{equation}

The representative radius is typically used in the calculation of settling velocity of the size distribution \citep[e.g.][]{Ackerman2001}, which we assume for the particle settling scheme in {\tt mini-haze}.
In the GCM, the two moments are advected in the atmosphere with the dynamical core and a vertical settling routine used to calculate the settling rate of the moments.
As in \citet{Lee2024b}, we apply mixing length theory to perform dry convective evolution of the temperature gradient in convective regions.
A vertical diffusion scheme using the resulting $K_{\rm zz}$ profile from mixing length theory is then used to transport tracers to mimic the effects of tracer transport from convective motions.
A minimum $K_{\rm zz}$ = 10$^{5}$\,cm$^{2}$\,s$^{-1}$ background eddy diffusion rate is applied following \citet{Ackerman2001} and \citet{Christie2022}.
The quantities that are evolved in the GCM are the particle number mixing ratio of the zeroth moment, $q_{0}$ [cm$^{3}$ cm$^{-3}$], and mass mixing ratio of the first moment, $q_{1}$ [g g$^{-1}$],  which are
\begin{equation}
    q_{0} = \frac{n_{\rm h}}{n_{\rm a}},
\end{equation}
and
\begin{equation}
    q_{1} = \frac{\rho_{\rm h}}{\rho_{\rm a}},
\end{equation}
respectively, where $n_{\rm a}$ [cm$^{-3}$] is the atmospheric number density.

\subsection{Particle-particle collisions}

Particle-particle collisional growth is the main evolutionary process for the haze particle size distribution in {\tt mini-haze}.
We include collisions through Brownian motion (coagulation) and differential gravitational collisions (coalescence) following the scheme developed by \citet{Lee2025a}.
The total collisional rate is given by the sum of the two processes
\begin{equation}
    \dot{L}_{\rm coll} = \left(\frac{dn_{\rm h}}{dt}\right)_{\rm coag} + \left(\frac{dn_{\rm h}}{dt}\right)_{\rm coal},
\end{equation}
which we briefly describe below.

In our two-moment haze model, following arguments by \citet{Rossow1978}, the collisional rate equations are derived assuming the size distribution follows a delta-peak (i.e. a monodisperse size distribution) and the dominant collisions are between same-sized particles \citep{Lee2025a}.
First, we define quantities that represent the local properties of the atmosphere and dynamical state of the haze particles.
$\eta_{\rm a}$ [g cm$^{-1}$ s$^{-1}$] is the atmospheric dynamical viscosity, given here by the \citet{Rosner2012} fitting function
\begin{equation}
\label{eq:rosner}
\eta_{\rm a} = \frac{5}{16}\frac{\left(\pi m k_{\rm b}T\right)^{1/2}}{\pi d^{2}}\frac{(k_{\rm b}T/\epsilon_{\rm LJ})^{0.16}}{1.22},
\end{equation}
where the parameters for the molecular diameter, $d$ [cm], mass, $m$ [g], and Lennard-Jones potential, $\epsilon_{\rm LJ}$, for \ce{H2}, He, and H, the main components of hydrogen-rich atmospheres are listed in \citet{Lee2023}, though in {\tt mini-haze} we include values taken from \citet{Rosner2012} to account for all gas species used in the {\tt mini-chem} chemical kinetics scheme \citep{Tsai2022}.
To mix the viscosity of different background gas species, we use the \citet{Davidson1993} mixing law, which takes into account the mixing ratio and momentum exchange between gas species.

The Knudsen number, $Kn$, is the ratio of the atmospheric mean free path, $\lambda_{\rm a}$ [cm], to particle size
\begin{equation}
    Kn = \frac{\lambda_{\rm a}}{r_{\rm h}},
\end{equation}
with the mean free path given by \citep{Jacobson2005}
\begin{equation}
    \lambda_{\rm a} = \frac{2\eta_{\rm a}}{\rho_{\rm a}}\left(\frac{\pi\bar{\mu}}{8RT}\right)^{1/2},
\end{equation}
where $\bar{\mu}$ [g mol$^{-1}$] is the local atmospheric mean molecular weight.
The settling velocity in the Stokes regime ($Kn$ $\ll$ 1), $v_{\rm f, Stokes}$ [cm s$^{-1}$], of the haze particles is given by the expression \citep{Ohno2018}
\begin{equation}
    v_{\rm f, Stokes} = \frac{2\beta gr_{\rm h}^{2}(\rho_{\rm d} - \rho_{\rm a})}{9\eta_{\rm a}}\left[1 + \left(\frac{0.45gr_{\rm h}^{3}\rho_{\rm a}\rho_{\rm d}}{54\eta_{\rm a}^{2}}\right)^{2/5}\right]^{-5/4},
\end{equation}
where we have reintroduced the contribution from buoyancy term ($\rho_{\rm d} - \rho_{\rm a}$), $g$ [cm s$^{-2}$] is the gravity, and $\beta$ the Cunningham slip factor given by \citep{Kim2005}
\begin{equation}
    \beta = 1 + Kn \left[1.165 + 0.483e^{-0.997/Kn}\right].
\end{equation}
However, in the $Kn$ $\gg$ 1 regime, which is commonly a regime where small particles such as haze in exoplanet atmospheres are found, the Epstein drag law should be used.
For this, we can follow \citet{Woitke2003} who used the \citet{Schaaf1963} drag coefficients in the limiting case for settling velocities well below the atmospheric thermal velocity, $v_{\rm f}$ $\ll$ $c_{\rm T}$,  to derive
\begin{equation}
    v_{\rm f, Epstein} = \frac{\left(\pi\right)^{1/2} g \rho_{\rm d} r_{\rm h}}{2\rho_{\rm a} c_{\rm T}},
\end{equation}
where $c_{\rm T}$ = $\left(2k_{\rm b}T/m_{\rm a}\right)^{1/2}$ [cm s$^{-1}$] is the thermal velocity of the atmosphere, with $m_{\rm a}$ [g] the mean mass of the atmosphere.

To interpolate between the Stokes and Epstein regimes, we use a simple tanh function to smoothly interpolate between the two limiting expressions
\begin{equation}
    f(Kn') = \frac{1}{2}[1 - \tanh(a\log Kn')],
\end{equation}
where $Kn'$ = $Kn$/$Kn$$_{\rm cr}$ with $Kn_{\rm cr}$ a critical transition Knudsen number.
The settling velocity is then the linear combination of this function between the limiting regimes
\begin{equation}
    v_{\rm f} =  f(Kn')v_{\rm f, Stokes} + \left[1 - f(Kn')\right]v_{\rm f, Epstein}.
\end{equation}
\citet{Woitke2003} found a critical value of $Kn_{\rm cr}$ = 1/3 using their derivations, but here we use the values from \citet{Lee2025b} who found that $Kn_{\rm cr}$ = 1 with a scaling factor of $a$ = 2 gives a good balance and smooth transition between the Stokes and Epstein regimes for the above tanh scheme.

Following the methods by \citet{Lee2025a}, in the two-moment monodisperse scheme, we include collisional growth through Brownian motion of the particles.
First, we define the particle diffusion factor, $D(r)$ [cm$^{2}$ s$^{-1}$], \citep{Chandrasekhar1943}
\begin{equation}
    D(r) = \frac{k_{\rm b}T\beta}{6\pi\eta_{\rm a}r},
\end{equation}
and particle thermal velocity, $V(m)$ [cm s$^{-1}$]
\begin{equation}
    V(m) = \left(\frac{8k_{\rm b}T}{\pi m}\right)^{1/2},
\end{equation}
where $m$ [g] is the mass of the particle.

In \citet{Moran2022}, a simple and accurate interpolation function between the continuum ($Kn$ $\ll$ 1) and free-molecular ($Kn$ $\gg$ 1) regime is theoretically derived.
This is a function of the diffusive Knudsen number, $Kn_{\rm D}$, which for a monodisperse size distribution is given as \citep{Moran2022}
\begin{equation}
    Kn_{\rm D} = \frac{8D(r_{\rm h})}{\pi V(m_{\rm h})r_{\rm h}}.
\end{equation}
The interpolation function, $g(Kn_{\rm D})$, derived by \citet{Moran2022}, is stated as
\begin{equation}
    g(Kn_{\rm D}) = \left(1 + \frac{\pi^{2}}{8} Kn_{D}^{2}\right)^{-1/2},
\end{equation}
which varies between a value of 1 for the continuum regime ($Kn_{\rm D} \rightarrow$ 0) and recovers the free molecular regime kernel when $Kn_{\rm D} \rightarrow$ $\infty$.
The rate of change of the haze particle number density from coagulation is then given as the monodisperse rate in the $Kn$ $\ll$ 1 regime \citep[e.g.][]{Lee2025a} modified by the \citet{Moran2022} interpolation function
\begin{equation}
\label{eq:N_coag_1_int}
    \left(\frac{dn_{\rm h}}{dt}\right)_{\rm coag} \approx - \frac{4k_{\rm b}T\beta}{3\eta_{\rm a}}n_{\rm h}^{2}g(Kn_{\rm D}).
\end{equation}

The gravitational coalescence term is given by \citep{Rossow1978, Ohno2018}  
\begin{equation}
    \left(\frac{dn_{\rm h}}{dt}\right)_{\rm coal} \approx -2\pi r_{\rm h}^{2}\epsilon v_{\rm f} E n_{\rm h}^{2},
\end{equation}
where $\epsilon$ is a parameter that estimates the relative velocity of the particles to the mean particle size settling velocity, $v_{\rm f}$ [cm s$^{-1}$] (i.e. $\Delta$$v_{\rm f}$ = $\epsilon$$v_{\rm f}$). 
This is taken as $\epsilon$ = 0.5 following the results of \citet{Sato2016}, who found this value to best reproduce the results of a collisional bin-resolving model for protoplanetary disk simulations.

$E$ is a collisional efficiency factor, dependent on the Stokes number, $Stk$, 
\begin{equation}
    Stk = \frac{\Delta v_{\rm f} v_{\rm f}(r_{\rm h})}{gr_{\rm h}} = \frac{\epsilon v_{\rm f}^{2}(r_{\rm h})}{gr_{\rm h}}.
\end{equation}
$E$ is then given by \citep{Guillot2014}
\begin{equation}
     E =
    \begin{cases}
      {\rm max}\left[0, 1 - 0.42Stk^{-0.75}\right], & Kn < 1 \\
      1, &  Kn \geq 1 
    \end{cases}
\end{equation}

The effect of particle collisions on the particle size distribution is to reduce the particle number density, as collisional growth is a sink term for the zeroth moment, which then increases the average radius of the distribution through Eq. \eqref{eq:mh} and Eq. \eqref{eq:rv}.
The first moment is not affected by collisions as the total mass is conserved in each collision.
This then physically accounts for the correct behaviour of collisional growth processes on the size distribution in the moment method, in line with bin resolving models \citep[e.g.][]{Gao2023}.

\subsection{Haze particle thermal decomposition}

For this study, we use a simple thermal loss timescale, $\tau_{\rm loss}$ [s], to estimate the thermal decomposition rate of haze particles.
For simplicity, we assume thermal decomposition occurs for haze particles that fall below a certain pressure level, $p_{\rm loss}$ [bar], \citep[e.g.][]{Steinrueck2023}.
For the zeroth moment this is given by
\begin{equation}
 \dot{L}_{\rm deco} =
    \begin{cases}
     -n_{\rm h}/\tau_{\rm loss}, & p \geq p_{\rm loss} \\
      0, &  p < p_{\rm loss}
    \end{cases}
\end{equation}
For the first moment the corresponding decomposition loss is $-\rho_{\rm h}$/$\tau_{\rm loss}$, equivalent to $m_{\rm h}\dot{L}_{\rm deco}$.
We follow \citet{Steinrueck2021, Steinrueck2023} and adopt a decomposition pressure of $p_{\rm loss}$ = 0.1\,bar in the GCM simulations.
This choice is in line with the results of \citet{Ohno2024} who explicitly calculated the decomposition of haze particles by oxidization and atomic hydrogen attack.

\subsection{Haze production scheme}
\label{sec:prod}

The photochemical conditions and chemical kinetic reaction network that give rise to formation of photochemical haze particles are complex and highly uncertain, containing many steps that build up from the initial precursor molecules \citep[e.g.][]{Pentsak2024}.
For example, the experimental results of \citet{Moran2020} that emulated the chemical conditions of a hydrogen-rich exoplanet atmosphere suggest a complex mixture of heavy oxygenated and nitrogenated hydrocarbon composition for the haze.
Due to this complexity and uncertainty, we have designed {\tt mini-haze} to be flexible with the origin of the precursor molecules and the mechanism that forms haze particles.
For the two-moment method, different production schemes provide additional source terms to both moment time evolution equations.
We initially include only a simple pressure dependent scheme below, leaving more complex formation mechanisms to future studies.

\subsection{Pressure-dependent formation rate}

A common way to parameterise the haze mass formation rate is to use a log-normal distribution in pressure around a median value representing the region of maximal production of precursor molecules.
This is typically given by a mass mixing ratio production rate, $\dot{P}_{\rm pre}$ [g g$^{-1}$ s$^{-1}$ $\equiv$ s$^{-1}$],  expression \citep[e.g.][]{Steinrueck2023}
\begin{equation}
\label{eq:p_pre}
    \dot{P}_{\rm pre} = \mu_{*}P_{0}\frac{g}{\left(2\pi\right)^{1/2}p\sigma}e^{-\frac{\ln^{2}(p/p_{\rm m})}{2\sigma^{2}}},
\end{equation}
where $\mu_{*}$ is the stellar zenith cosine angle, $P_{0}$ [g cm$^{-2}$ s$^{-1}$] is the column integrated mass production rate of the haze particle precursor species at the substellar point, $g$ [cm s$^{-2}$] the atmospheric gravity, $p$ [dyne cm$^{-2}$] the local pressure, $\sigma$ the standard deviation and $p_{\rm m}$ [dyne cm$^{-2}$] the median pressure value of haze precursor production.
It is important to note that, in this study, we use the above equation to represent the formation rate of haze particle precursor species, rather than the solid haze particle formation rate directly as by \citet{Steinrueck2021, Steinrueck2023}.
For example, a conversion rate of $\approx$ 1\% for a precursor production rate of 2.5 $\cdot$ 10$^{-10}$\,g\,cm$^{-2}$\,s$^{-1}$ would be similar to a haze particle production rate of 2.5 $\cdot$ 10$^{-12}$\,g\,cm$^{-2}$\,s$^{-1}$, assuming instantaneous conversion.
For comparison, \citet{Lavvas2010} used a column mass production rate of $\sim$ 3.0 $\cdot$ 10$^{-14}$\,g\,cm$^{-2}$\,s$^{-1}$ for Titan haze modelling, two orders of magnitude less than that used here.
 
\subsection{Delayed formation rate}

Recently, \citet{Tsai2023} presented 2D photochemical kinetics modelling of WASP-39b, suggesting the dynamical transport of photochemical products and radicals such as S and OH from their dayside formation regions to the nightside, which then recombine into larger species such as SO$_{2}$.
This mechanism changes the limb-to-limb chemical structure and transmission spectra when compared to a 1D modelling approach \citep[e.g.][]{Tsai2023a}.
In addition, \citet{Powell2024} presented a pseudo-2D version of the {\tt CARMA} microphysical condensate cloud model applied to a set of hot Jupiter models, finding that the global cloud structure strongly depends on the dynamical formation of cloud species on the nightside regions of the planet, which are then advected across into the dayside.

To emulate a similar nightside formation process, we propose a `delayed formation timescale' for the haze particles from their initial gas phase precursor species.
This has been examined in a similar manner by \citet{Bertrand2017}, who assumed an exponential formation time of $\tau$ $\approx$ 10$^{7}$\,s for their precursor material in Pluto haze GCM simulations, and also performed some sensitivity studies for the parameter.

Inspired by the results of \citet{Tsai2023} and \citet{Powell2024}, we develop a simple timescale-based parameterisation, where haze precursor molecules form on the dayside, but the end state formation of the solid haze particles can be delayed. 
Through the balance of timescale choices, we can examine, for example, a scenario where precursors form on the dayside of the atmosphere, but the haze formation process itself is not confined to the dayside of the planet.

We propose a coupled set of chemical timescale equations, where we differentiate between a generic inactive precursor, $q_{\rm pre}$, and activated precursor tracer, $q_{\rm act}$, that is then able to go on and form haze monomers.
To describe the evolution of the generic precursor tracer we use
\begin{equation}
    \frac{dq_{\rm pre}}{dt} = \dot{P}_{\rm pre} - \frac{q_{\rm pre}}{\tau_{\rm act}} - \frac{q_{\rm pre}}{\tau_{\rm d, pre}},
\end{equation}
where $\dot{P}_{\rm pre}$ [g g$^{-1}$ s$^{-1}$] is the production rate mass mixing ratio of precursor molecules, $\tau_{\rm act}$ [s] the precursor activation chemical timescale and $\tau_{\rm d, pre}$ [s] the precursor species decay timescale, representing an exponential-in-time decrease where the precursor species evolves into non-precursor species.
In this study, we use a pressure-dependent, parameterised $\dot{P}_{\rm pre}$ from Eq. \eqref{eq:p_pre}, however, different production rate schemes can be used, such as coupling a precursor rate from a chemical network.
The evolution of activated species follows
\begin{equation}
    \frac{dq_{\rm act}}{dt} = \frac{q_{\rm pre}}{\tau_{\rm act}} - \frac{q_{\rm act}}{\tau_{\rm form}} - \frac{q_{\rm act}}{\tau_{\rm d, act}},
\end{equation}
where again a decay timescale, $\tau_{\rm d, act}$ [s], is given, representing an exponential-in-time decay of activated precursor species.
In this study, we assume the same decay timescale for the precursors and activated species, but this can be readily altered in future studies where chemical models can inform on how the timescales differ, which may be temperature and pressure dependent.
The haze monomer formation rate, $\dot{P}_{\rm h}$ [g g$^{-1}$ s$^{-1}$ $\equiv$ s$^{-1}$], is then simply 
\begin{equation}
    \dot{P}_{\rm h} = \frac{q_{\rm act}}{\tau_{\rm form}},
\end{equation}
where $\tau_{\rm form}$ [s] is the timescale required to convert activated precursors into the solid phase.

In our framework, the net efficiency of the conversion of precursor material to solid haze particles is set through the relative ratios between the formation, activation and decay timescales of the system.
In a steady state, the relation between $\dot{P}_{\rm h}$ and $\dot{P}_{\rm pre}$ is given as
\begin{equation}
    \dot{P}_{\rm h}=\dot{P}_{\rm pre}\frac{1}{(1+\tau_{\rm act}/\tau_{\rm d,pre})(1+\tau_{\rm form}/\tau_{\rm d,act})}
\end{equation}
For example, to activate approximately 1\% of the precursor material before precursor decay, we require $\tau_{\rm act}$ $\approx$ 100 $\tau_{\rm d, pre}$, assuming the formation stage is efficient.
To recover a near-instantaneous conversion between precursor molecules and haze particles, the activation and formation timescales can be set to a small value such as $\sim$ 1--10\,s at the required efficiency ratio.
To delay a portion of the haze formation chain process, the activation timescale can be set to around a quarter to a half the advective timescale with a short formation timescale.
The decay timescale can then be tuned to only allow a certain fraction of the precursor material to survive the transport from the dayside to the nightside.
In practice, for this system, an equilibrium should eventually form between the decay of precursors and the rate of precursor formation across the global domain, which would then be in balance with the rate of the haze formation sequence timescales.

\section{Test application}
\label{sec:GCM}

\subsection{HD 189733b GCM simulation}

\begin{table*}
\centering
\caption{Adopted {\tt Exo-FMS} simulation parameters for the hazy HD 189733b hot Jupiter scenario.}
\begin{tabular}{l c l l}  \hline \hline
 Parameter & Value  & Unit & Description \\ \hline
 $T_{\rm int}$ & 390 & K & Internal temperature \\
 $p_{\rm 0}$ & 220 &  bar & Reference surface pressure \\
 $p_{\rm up}$ & 10$^{-7}$ &  bar & Upper boundary pressure \\
 $c_{\rm p}$ & 11334 &  J K$^{-1}$ kg$^{-1}$ & Specific heat capacity \\
 $R_{\rm d}$ & 3149 &  J K$^{-1}$ kg$^{-1}$  & Specific gas constant \\
 $\kappa$ & 0.28  & ...  & Adiabatic coefficient \\
 $g_{\rm p}$ & 23.1  & m s$^{-2}$ & Acceleration from gravity \\
 $R_{\rm p}$ &  8 $\cdot$ 10$^{7}$ & m & Planetary radius\\
 $\Omega_{\rm p}$ & 3.278 $\cdot$ 10$^{-5}$ & rad s$^{-1}$ & Rotation rate \\
 $\left[{\rm M/H}\right]$ & 1.0 & dex & solar metallicity \\
 $\tau_{\rm drag}$ & 10$^{5}$ & s & Basal drag timescale \\
 $\Delta$ t$_{\rm dyn}$ & 15 & s & Hydrodynamic time-step \\
 $\Delta$ t$_{\rm rad}$  & 15 & s & Radiative time-step \\
 $\Delta$ t$_{\rm chem}$ &  60 & s & {\tt mini-chem} time-step \\
 $\Delta$ t$_{\rm haze}$ &  60 & s & {\tt mini-haze} time-step \\
 $\Delta$ t$_{\rm MLT}$ &  0.5 & s & mixing length theory time-step \\
 $P_{0}$ & 2.5 $\cdot$ 10$^{-9}$ & kg m$^{-2}$ s$^{-1}$ & Haze precursor production rate \\
 $p_{\rm m}$ & 2 $\cdot$ 10$^{-6}$ & bar & Median production pressure \\
 $\sigma$ & 0.576 & ... & Standard deviation of production zone \\
 $\tau_{\rm act}$ & 10:56000 & s & Precursor activation timescale \\
 $\tau_{\rm form}$ & 10:100 & s & Haze formation timescale \\
 $\tau_{\rm d, pre}$ & 1:560 & s & Precursor decay timescale \\
 $\tau_{\rm d, act}$ & 1:560 & s & Activated species decay timescale \\
 $\tau_{\rm loss}$ & 1000  & s & Thermal loss timescale \\
 $p_{\rm loss}$ & 0.1  & bar & Thermal loss boundary \\
 $N_{\rm v}$ & 56  & ... & Number of vertical layers in GCM \\
\hline
\end{tabular}
\tablefoot{We use a cubed-sphere resolution of C32 ($\approx$  128 $\times$ 64 in longitude $\times$ latitude).
We adopt and derive system parameters following the values published by \citet{Addison2019} and follow a similar setup and haze parameters of \citet{Steinrueck2023}.}
\label{tab:GCM_parameters}
\end{table*}

In this initial study, we couple {\tt mini-haze} to the {\tt Exo-FMS} GCM \citep[e.g.][]{Lee2021} and simulate a hazy hot Jupiter HD 189733b scenario, using a similar setup to \citet{Steinrueck2023}.
Haze formation in the atmosphere of HD 189733b has been examined in previous 1D studies \citep[e.g.][]{Lavvas2017} and invoked as a possible mechanism to explain the strong Rayleigh slope-like feature seen in HST transmission spectra data \citep[e.g.][]{Pont2013, Sing2016}.
Our HD 189733b GCM study presents a useful test case in order to examine how the moment approach, with time dependent haze evolution, affects the global atmospheric structures and leads to vertical and horizontal haze particle inhomogeneities in haze particle mass mixing ratio and particle sizes.

To integrate the ODE system for the moments and precursor tracers, we use the implicit, stiff ODE solver {\tt DLSODE}\footnote{\url{https://computing.llnl.gov/projects/odepack/software}} \citep{Radhakrishnan1993}. 
Overall, four tracers are required to be evolved in the GCM simulation to couple {\tt mini-haze}, the volume mixing ratio of the zeroth moment, the mass mixing ratio of the first moment, the mass mixing ratio of precursor molecules and the mass mixing ratio of activated precursor molecules. 
For our HD 189733b simulations, we assume a haze monomer radius of $r_{0}$ = 1\,nm and a soot-like bulk density of $\rho_{\rm d}$ = 1\,g\,cm$^{-3}$.
We find that small chemical and haze time steps (60\,s), and dynamical and radiative time steps on the order of 15\,s, are required to keep the simulation stable, similar to that found in the \citet{Steinrueck2023} simulations.

Recent observations of HD 189733b suggest a super-solar metallicity atmosphere for HD 189733b.
From analysis of JWST NIRCam transmission spectra data, \citet{Fu2024} suggested an atmospheric metallicity of 2--5$\times$ solar.
From analysis of high-resolution $K$-band emission spectra, \citet{Finnerty2024} suggested a super-solar C/H and O/H ratio. 
We therefore use the $\left[{\rm M/H}\right]$ = 1.0\,dex set of net forward chemical rates from {\tt mini-chem} \citep{Tsai2022, Lee2023}.
This may overestimate the metallicity of the atmosphere, but is the closest parameter match in the current {\tt mini-chem} chemical database.

A significant difference between the \citet{Steinrueck2021,Steinrueck2023} methodology and the present work is the value of the haze particle formation rates.
In \citet{Steinrueck2021,Steinrueck2023} the variable $F_{0}$ [kg m$^{-2}$ s$^{-1}$] represents directly the haze particle formation rate.
However, instead of the \citet{Steinrueck2021,Steinrueck2023} scheme, in the current study we parameterise the formation of haze particle precursor species using the mass mixing ratio production variable, $\dot{P}_{\rm pre}$, which represents the gas phase precursor production rate, rather than direct solid haze particle formation.
This additional step attempts to broadly emulate a photochemical haze formation process driven by an initial reservoir of precursor gas materials \citep[e.g.][]{Morley2015} that goes through additional chemical processing, which eventually leads to some fraction of the precursor material being incorporated into the solid haze product.

\subsection{Haze opacity feedback}

We use the same scheme as by \citet{Lee2025a} to perform the radiative feedback of the haze particles onto the atmospheric temperature-pressure structure inside the GCM simulations.
In summary, for small size parameters ($x$ $<$ 0.01) we apply the Rayleigh limit expressions \citep{Bohren1983} and for large size parameters ($x$ $>$ 10) we use modified anomalous diffraction theory following \citet{Moosmuler2018}.
For intermediate size parameters, we use the {\tt LX-MIE} code from \citet{Kitzmann2018}.
This scheme avoids using Mie theory for large size parameters which can be computationally expensive, while retaining good accuracy that captures the salient effects of haze opacity feedback.
We include the calculation of the single scattering albedo and asymmetry factor within the Mie theory calculation, passing these values through the GCM $n$-stream multiple scattering radiative-transfer routine based on \citet{Toon_1989}.

For the input real, $n$, and imaginary, $k$, optical constants, we assume carbon soot particles, taking data from \citet{Lavvas2017}.
However, {\tt mini-haze} can use any tabulated literature optical constants, for example Titan tholins \citep[e.g.][]{Khare_1984, He_2022} and exoplanet atmospheric condition hazes \citep[e.g.][]{Corrales2023, He2024}.

\subsection{Coupling {\tt Exo-FMS} and {\tt mini-haze}}
\label{sec:res}

\begin{figure*}
    \centering
    \includegraphics[width=0.49\linewidth]{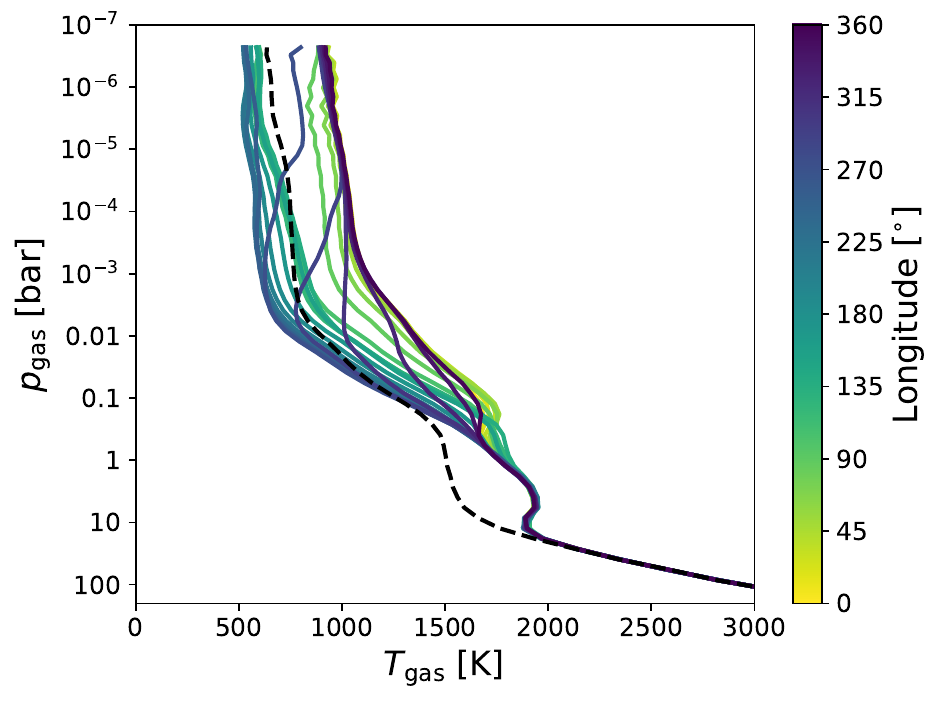}
    \includegraphics[width=0.49\linewidth]{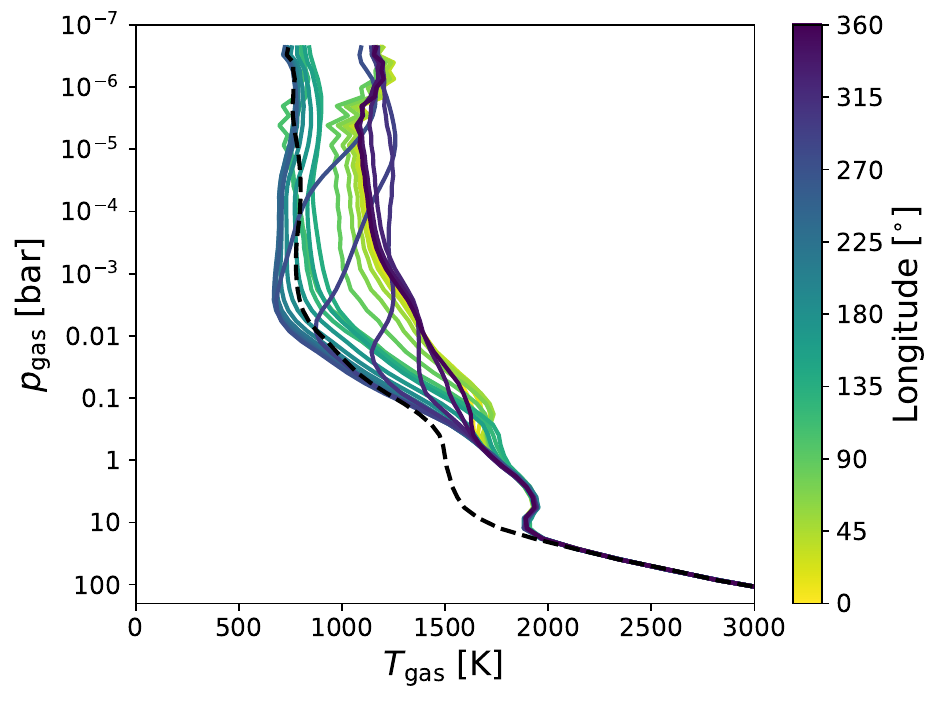}
    \includegraphics[width=0.49\linewidth]{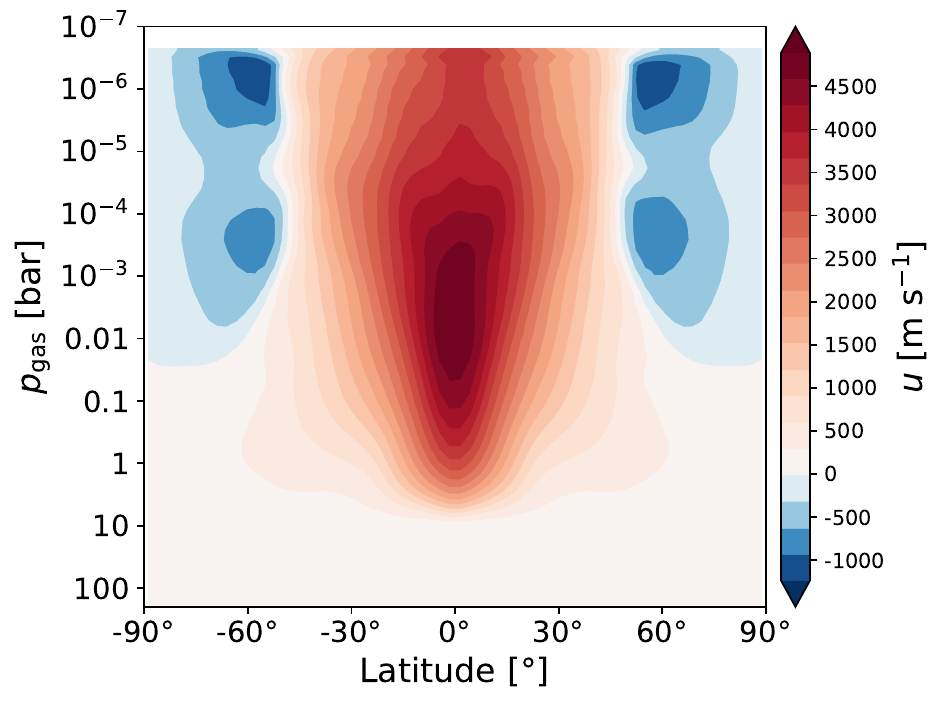}
    \includegraphics[width=0.49\linewidth]{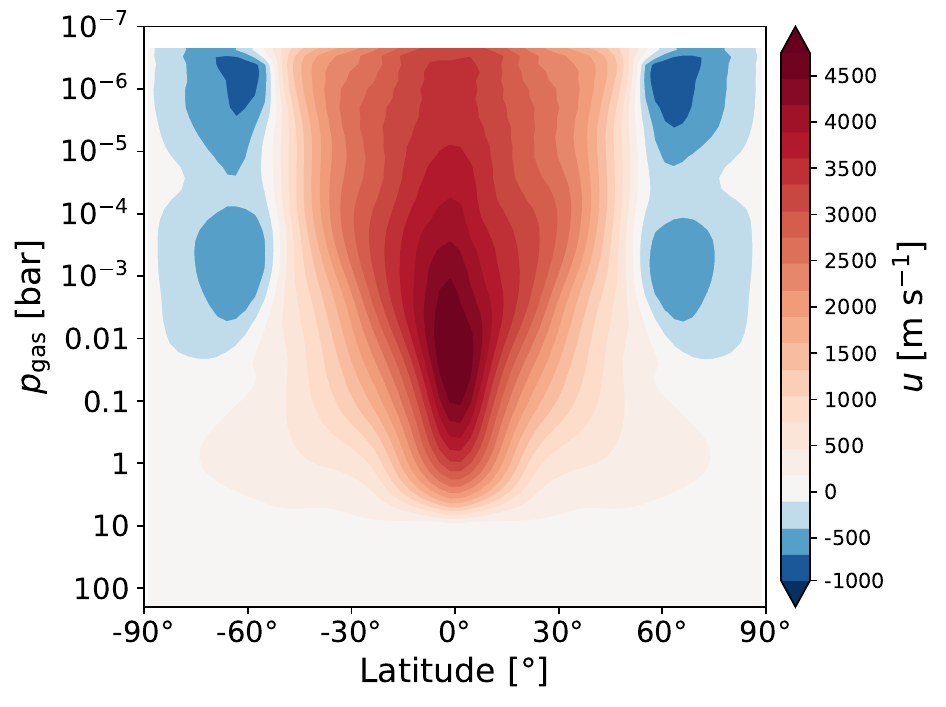}
    \caption{Temperature-pressure ($T$-$p$) profiles (top row) at the equatorial region as a function of longitude (colourbar) for the short (left) and long (right) timescale formation simulation. 
    The dashed line shows a polar $T$-$p$ profile.
    Zonal mean zonal velocity (bottom row) of the HD 189733b simulation for the short (left) and long (right) timescale formation simulation.}
    \label{fig:gcms}
\end{figure*}

In our first GCM simulation of a hazy HD 189733b scenario, we investigate a near instantaneous formation timescale of haze particles where precursors are formed. 
We assume an activation timescale, $\tau_{\rm act}$ = 10\,s, a formation timescale of $\tau_{\rm form}$ = 10\,s and a decay timescale of $\tau_{\rm decay}$ = 1\,s.
This enables a quick transformation between the precursors and haze particles at a modest $\approx$1\% efficiency rate.
This simulation is given the label `short timescale formation'.

As opposed to the first GCM simulation, we assume a delayed haze formation sequence in our second simulation.
For this, we assume that the activation timescale is equal to one half the advective timescale of the atmosphere, to ensure a portion of the overall haze formation regions occur on the nightside hemisphere of the planet.
From the GCM results, the zonal mean velocity in the upper atmosphere is $u$ $\sim$ 4500\,m s$^{-1}$, which results in an advective timescale of $\tau_{\rm ad}$ $\sim$ 1.12 $\cdot$ 10$^{5}$\,s.
This leads to an activation timescale, $\tau_{\rm act}$ = 56000\,s and we assume a formation timescale of $\tau_{\rm form}$ = 100\,s.
We increase the decay timescale to $\tau_{\rm decay}$ = 560\,s, to ensure an overall net 1\% of precursor material forms the haze particles.
Since $\tau_{\rm form}$ $<$ $\tau_{\rm decay}$ in this case, activated material can easily form with minimal loss, mimicking an efficient haze formation zone once precursor material enters the nightside.
In our scheme, the transport of materials to the nightside therefore also naturally reduces the activation efficiency as more materials decay during the time it takes to transport to the nightside, rather than form directly at the precursor production sites.
This simulation is given the label `long timescale formation'.

We include non-equilibrium chemistry at 10$\times$ solar metallicity using the {\tt mini-chem} miniature thermokinetic network scheme \citep{Tsai2022,Lee2023}, and take into account the changing chemical composition and subsequent gas phase opacity into the radiative-transfer model using the adaptive equivalent extinction method presented by \citet{Amundsen2017}.
We use a correlated-k scheme with 11 bands the same as \citet{Kataria2013} and produce k-tables for the {\tt mini-chem} species: OH \citep{Hargreaves2019}, \ce{H2O} \citep{Polyansky2018}, CO \citep{Li2015}, \ce{CO2} \citep{Yurchenko2020}, \ce{CH4} \citep{Hargreaves2020}, \ce{C2H2} \citep{Chubb2020}, \ce{NH3} \citep{Coles2019}, and HCN \citep{Harris2006}, where the citation for each species is the source of line-list data used to produce the gas phase opacities for that species.
We include collision-induced absorption opacity of \ce{H2}-\ce{H2}, \ce{H2}-He, \ce{H2}-H, and He-H collisional pairs using the HITRAN database \citep{Karman2019} and include Rayleigh scattering opacity from \ce{H2}, He, and H. 
In addition, we include the gas phase opacity of Na and K, assuming they are quenched at constant volume mixing ratio values of 10$^{-5}$ and 10$^{-6}$ respectively, which is approximately 10$\times$ their solar abundance \citep{Woitke2018}.

We perform both simulations for 2000 Earth days without radiative feedback from the haze particles, but including the dynamical evolution of the haze, after which we include haze opacity for another 500 days.
This spin-up strategy is chosen to avoid the very small dynamical and radiative timesteps, $\sim$ 15\,s, required by the GCM to remain stable when radiative-feedback of the haze is turned on.
Our simulations more quickly reach a statistically stable state using dynamical and radiative timesteps of $\sim$ 60\,s, including the global haze physical processes, before the thermal feedback slows down the computation significantly.
We take the averaged values across the final 100 days as the final result.

In Fig. \ref{fig:gcms}, we present the temperature-pressure ($T$-$p$) profiles for each simulation at the equatorial region and a polar profile.
Our results show that the $T$-$p$ profiles are altered significantly between the short and long timescale formation assumptions.
For the long timescale case, at very low pressures ($p$ $<$ 10$^{-5}$\,bar), a temperature inversion is formed, which persists onto the nightside of the planet, while for the short timescale simulation the haze opacity has less impact.
The mid-atmosphere shows a more isothermal structure leading up to the inverted upper atmosphere, which is most clear in the long timescale formation case.
The effects on the $T$-$p$ profiles are similar to that found in the soot particle cases by \citet{Steinrueck2023}, who also found an increase in the upper atmosphere temperatures, inversions on the dayside and nightside, as well as the more isothermal mid-atmosphere structure due to haze particle radiative feedback.
Overall, our results suggest that the delayed formation of haze has an overall larger radiative-feedback effect on the global $T$-$p$ structures compared to instantaneous formation.

Figure \ref{fig:gcms} also shows the zonal mean zonal velocity plot for the short and long formation timescale model. 
This shows a typical hot Jupiter zonal wind structure in the atmosphere, with a strong equatorial jet being the dominant dynamical driver of the system.
The overall mean structures of the jet and wind speeds are not altered significantly between the two assumptions.

\subsection{Haze production rate}

\begin{figure*}
    \centering
    \includegraphics[width=0.49\linewidth]{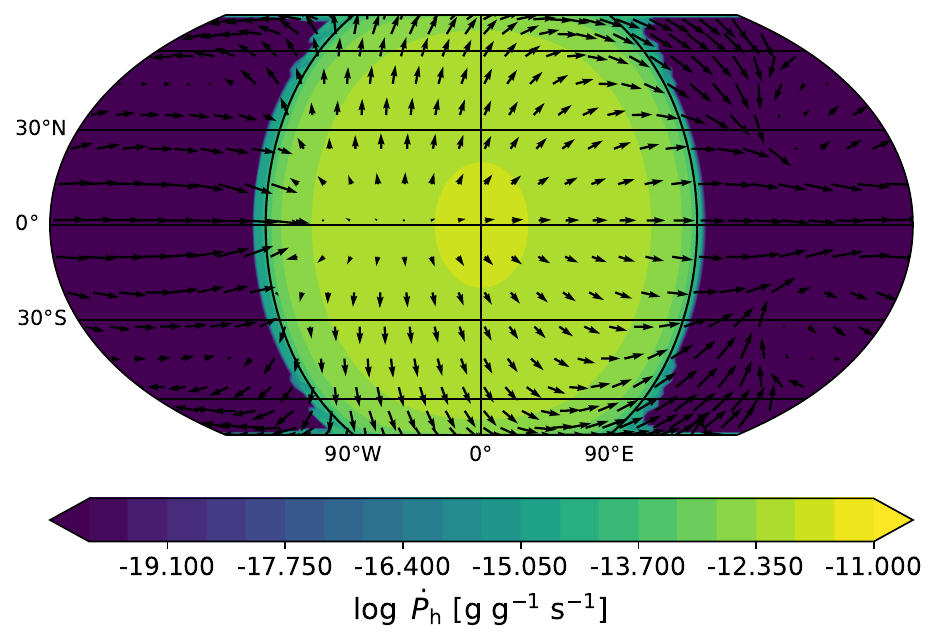}
     \includegraphics[width=0.49\linewidth]{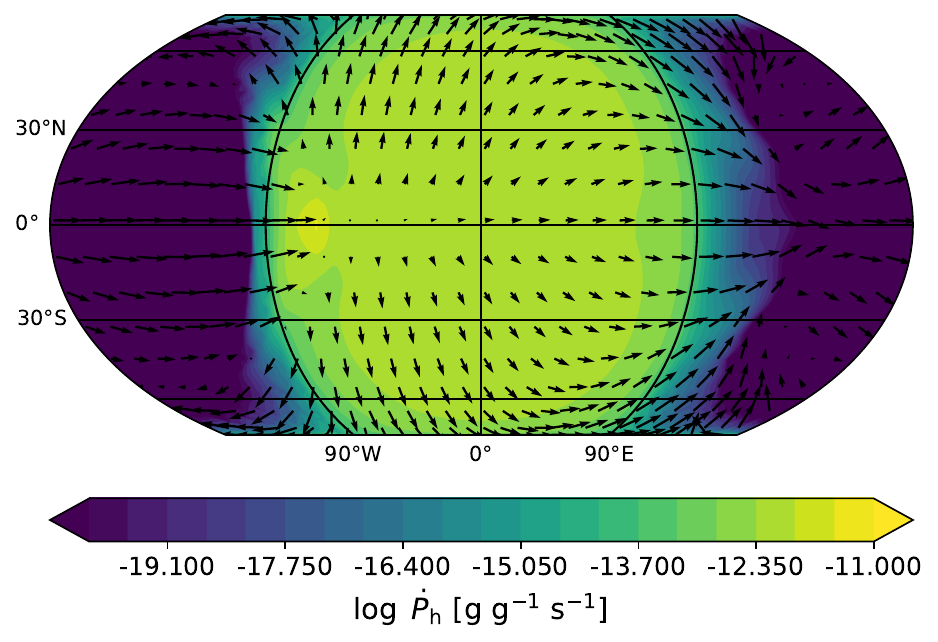}
    \caption{2D latitude-longitude map of the solid haze particle mass mixing ratio formation rate, $\dot{P}_{\rm h}$ [g g$^{-1}$ s$^{-1}$ $\equiv$ s$^{-1}$], at the 10$^{-5}$\,bar pressure level for the short formation timescale (left) and long formation timescale (right) GCM simulation.}
    \label{fig:Ph}
\end{figure*}

\begin{figure*}
    \centering
    \includegraphics[width=0.49\linewidth]{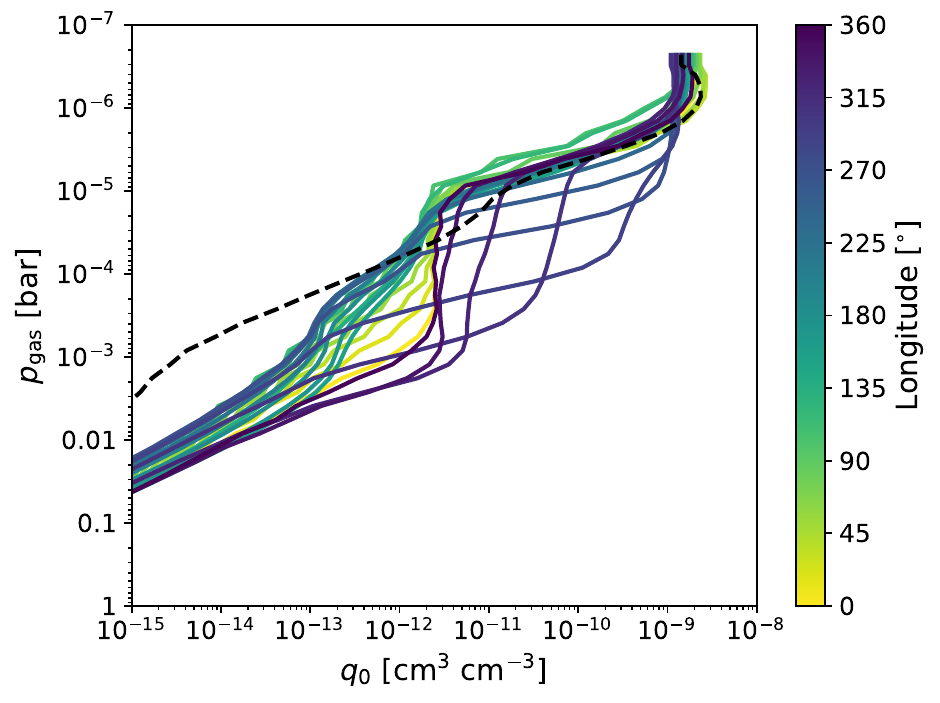}
    \includegraphics[width=0.49\linewidth]{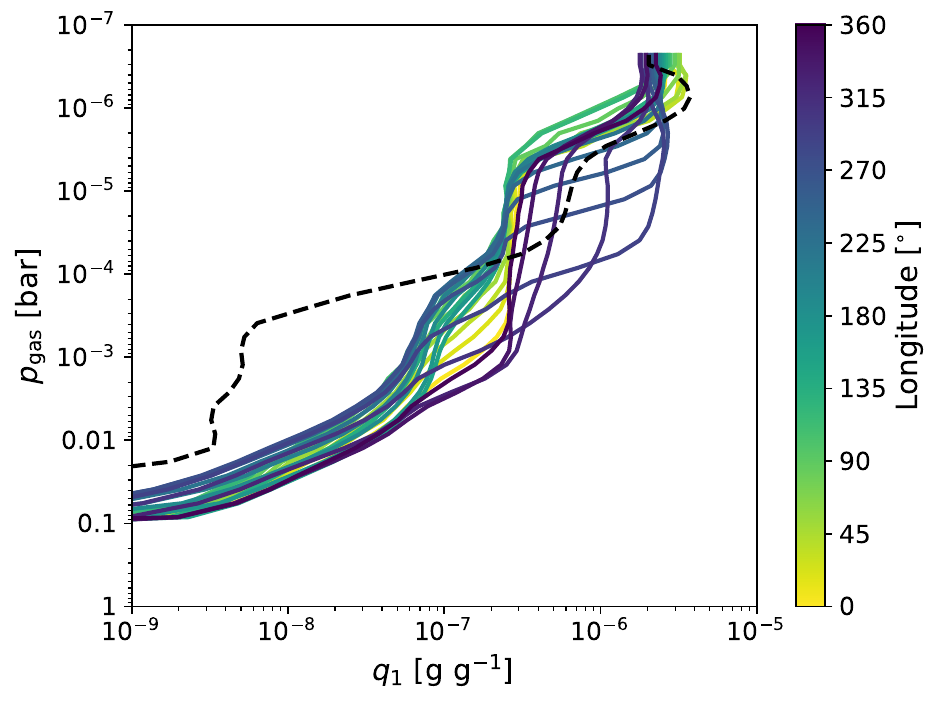}
    \includegraphics[width=0.49\linewidth]{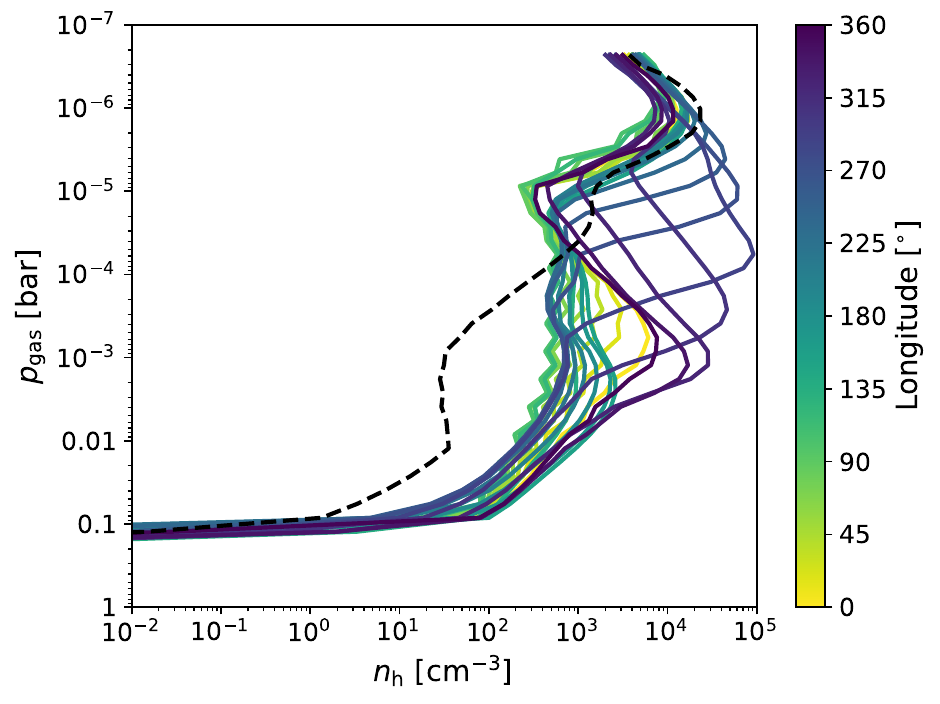}
    \includegraphics[width=0.49\linewidth]{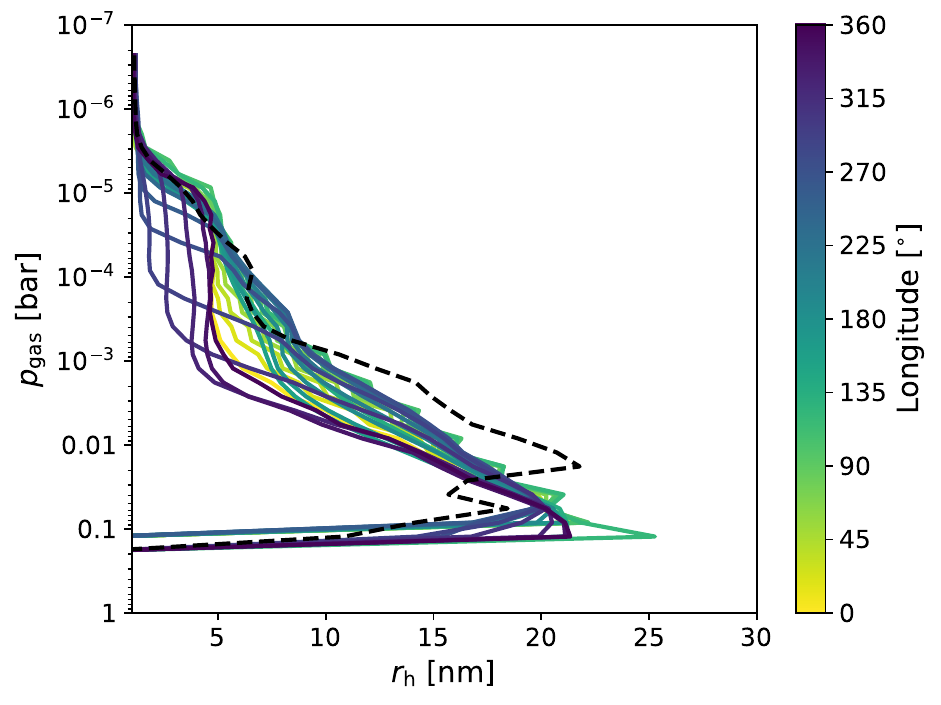}
    \caption{Short formation timescale vertical profiles of the zeroth moment volume mixing ratio, $q_{0}$ [cm$^{3}$ cm$^{-3}$] (top left), first moment mass mixing ratio, $q_{1}$ [g g$^{-1}$] (top right), haze particle number density, $n_{\rm h}$ [cm$^{-3}$] (bottom left) and haze particle radius, $r_{\rm h}$ [nm] (bottom right).
    Coloured lines denote the longitude (colour bar) at the equatorial region, while the black dashed line denotes a polar region.}
    \label{fig:vert_short}
\end{figure*}

In Fig. \ref{fig:Ph}, we show the solid haze formation rate, $\dot{P}_{\rm h}$ [g g$^{-1}$ s$^{-1}$] at the 10$^{-5}$\,bar pressure level for each GCM simulation.
Here the effect of the delayed timescale scheme is apparent, with the short timescale assumption confining the haze formation to the dayside of the planet, while the delayed scheme allows a significant portion of the haze particles to form on the nightside hemisphere.
The maximum regions of haze formation are also changed, with the short timescale scheme producing maximum haze near the sub-stellar point, while the long-timescale haze has its maximum near the western terminator. 
This is probably due to the western terminator downwelling region advecting precursor material to this pressure layer before it is converted to solid haze.

\subsection{Short formation timescale}

In Fig. \ref{fig:vert_short}, we present vertical profiles of the moment mixing ratios, haze number density and particle size for the short formation timescale simulation.
Our mass mixing ratio results show a similar structure to \citet{Steinrueck2023}, with values starting from the initial mixing ratios at the formation regions and decreasing with increasing pressure.
Our haze formation scheme produces a significant number density of $n_{\rm h}$ $>$ 10$^{2}$\,cm$^{-3}$ at pressures $p$ $<$ 0.1\,bar, but small particle sizes of a maximum of $\sim$ 20\,nm near the parameterised thermal decomposition pressure.

Figure \ref{fig:map_short} shows the 2D latitude-longitude maps of the haze mass mixing ratio and particle sizes at various isobar pressures in the simulation.
This shows a very similar 3D global distribution of haze particles to \citet{Steinrueck2023}, where the mass mixing ratio of the particles follows closely the vertical velocity structure of the atmosphere, with significant regions of higher mass mixing ratio at the western limb of the planet where downwelling from higher altitude occurs \citep{Steinrueck2023}.
The particle sizes follow a similar dynamical pattern to the mass mixing ratio, with larger particles generally present in regions of larger mixing ratio and vice versa.
This behaviour lines up with Fig. \ref{fig:vert_short}, where the number density and particle sizes follow an inverse pattern between each other with height.
However, this is not always the case, particularly the upper atmosphere ($p$ $<$ 10$^{-4}$\,bar), where larger mass mixing ratios are correlated with smaller particles.
At very low pressures, the collisional timescale is longer and the particle size is influenced more by whether the local flow is upwelling or downwelling.
In regions of downwelling, smaller particles from lower pressures are being mixed downwards, thus the particle size decreases. 
In regions of upwelling, larger particles from higher pressures are being transported upwards and the average particle size increases.
In contrast, deeper in the atmosphere, the collisional timescale is shorter and in regions of enhanced mass mixing ratio, particles will grow faster, leading to a greater correlation between mass mixing ratio and particle size.
Overall, our results suggest that the collisional behaviour and overall global structure of the haze particle is a result of a complex interaction of the flow patterns, settling rates and local collisional rates.

\subsection{Long formation timescale}

\begin{figure*}
    \centering
    \includegraphics[width=0.49\linewidth]{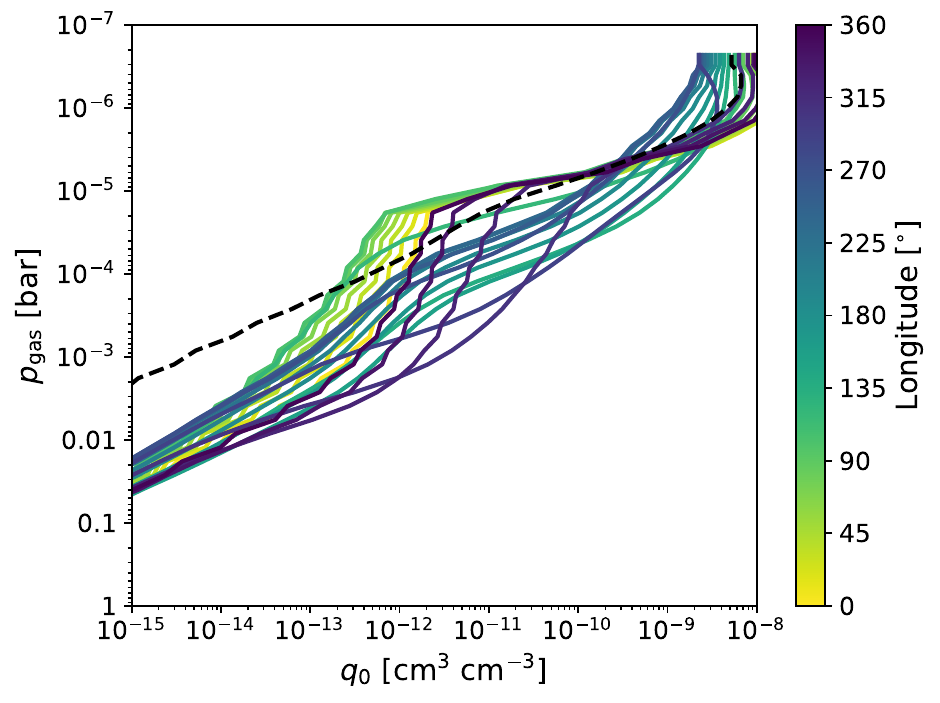}
    \includegraphics[width=0.49\linewidth]{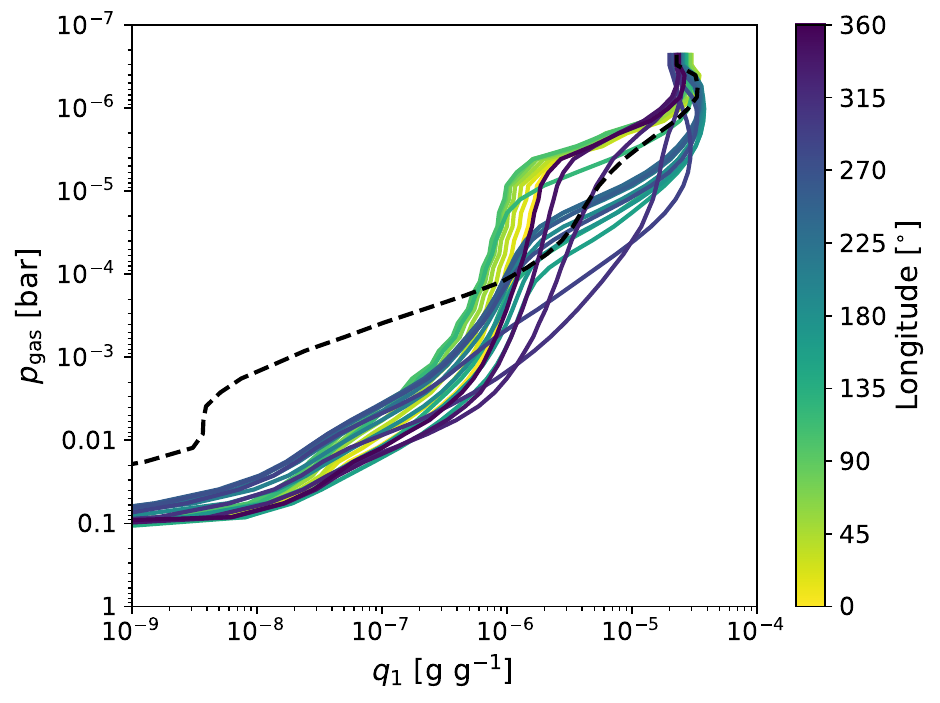}
    \includegraphics[width=0.49\linewidth]{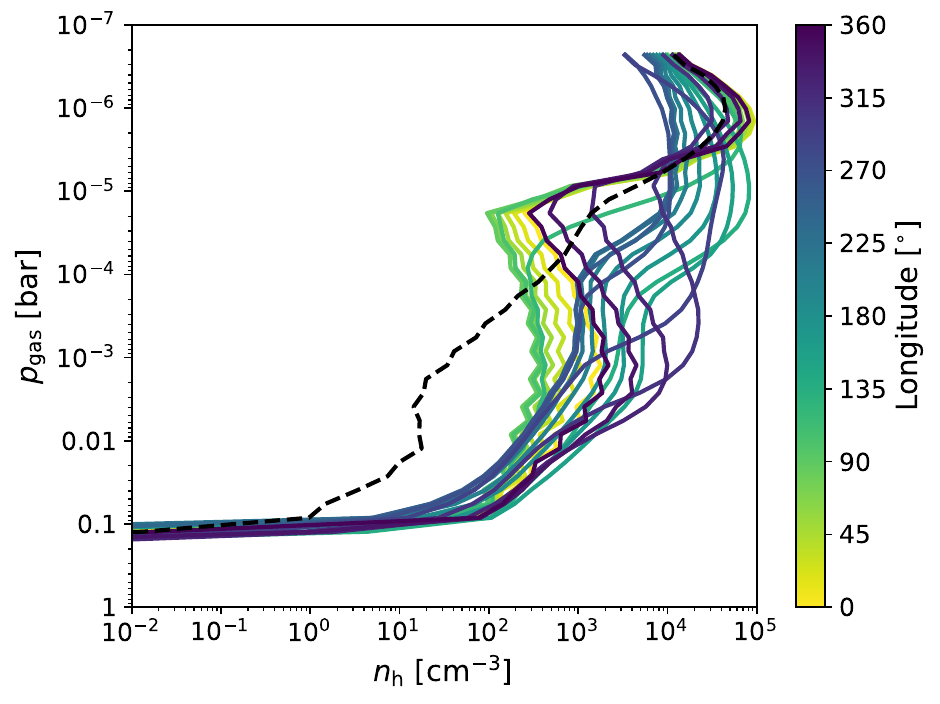}
    \includegraphics[width=0.49\linewidth]{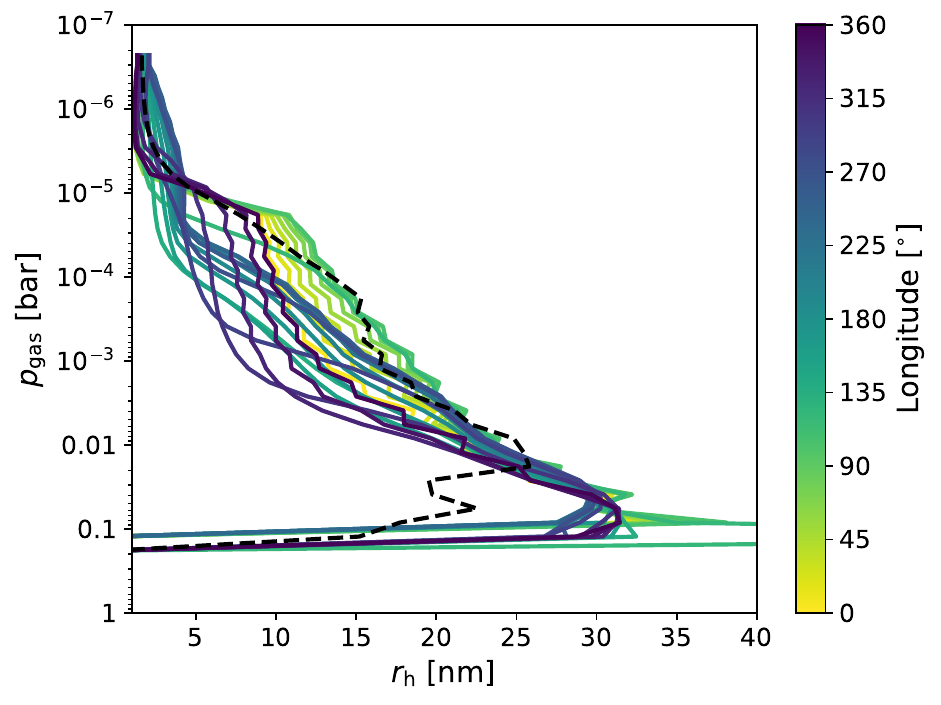}
    \caption{Same as Fig. \ref{fig:vert_short} but for the long haze formation timescale.}
    \label{fig:vert_long}
\end{figure*}

In Fig. \ref{fig:vert_long}, we present vertical profiles of the moment mixing ratios, haze number density and particle size for the long formation timescale simulation.
Here, similar profiles to the short timescale case are seen, but with enhancements of mixing ratios at the nightside of the planet, especially at the western limb regions compared to the short timescale structures.
The overall number density profiles are similar between the two cases, but the longer formation timescale produces larger particles, up to $\sim$ 30\,nm in the atmosphere compared to the short timescale $\sim$ 20\,nm.

Figure \ref{fig:map_long} shows the latitude-longitude maps of the haze mass mixing ratio and particle sizes at various isobar pressures in the simulation.
These show a highly similar pattern to the short timescale results, but with generally more enhancement of particles on the nightside hemisphere, and typically larger particles at each isobar pressure level.

Overall, the long timescale results suggest a generally faster coagulation rate compared to the short timescale simulations and a longer residence time allowing the haze to grow larger before settling out.
For the faster coagulation explanation, the increased particle sizes are likely due to an increase in the overall mass mixing ratios at the dynamically convergent east and west limb regions of the atmosphere, which leads to a larger coagulation rate and therefore larger particles.
For the longer residence explanation, the delayed formation timescale scheme naturally allows the smallest particles to spread across the globe before colliding and growing, retaining more material before they are dynamically converged and allowed to grow.
However, both mechanisms are not fast enough to overcome the natural dynamical timescales of the planet and decouple strongly from the flow, with haze particles still in the nanometre regime.

Our long timescale simulations produce an equatorial banding pattern of the largest particles ($\approx$ 30\,nm), similar to the banding of mass mixing ratio seen by \citet{Steinrueck2021} at 10$^{-2}$\,bar.
From our GCMs and the results of \citet{Steinrueck2021}, this region dynamically converges the mixing ratio of the haze particles, which in our model enables a larger particle size to be produced.
This banding is also seen in the short timescale simulations, but to a lesser extent, suggesting a link between the settling rate of particles and the dynamical structures at this pressure.
This is confirmed by \citet{Steinrueck2021}, who found larger particles increased the overall mass mixing ratio in this equatorial band compared to smaller particles.

In addition, the correlation between the mass mixing ratio and particle size occurs to a much greater extent at 10$^{-2}$\,bar compared to the lower pressure regions in both the short and longer timescale simulations, especially at the equatorial regions.
This suggests a dynamical convergence of both moment quantities when haze enters this dynamical layer.
This is also seen by \citet{Steinrueck2021} and \citet{Mak2025} who also found a strong convergence of the mass mixing ratio of haze at these pressure levels.

\subsection{Post-processing}
\label{sec:pp}

\begin{figure*}
    \centering
    \includegraphics[width=0.49\linewidth]{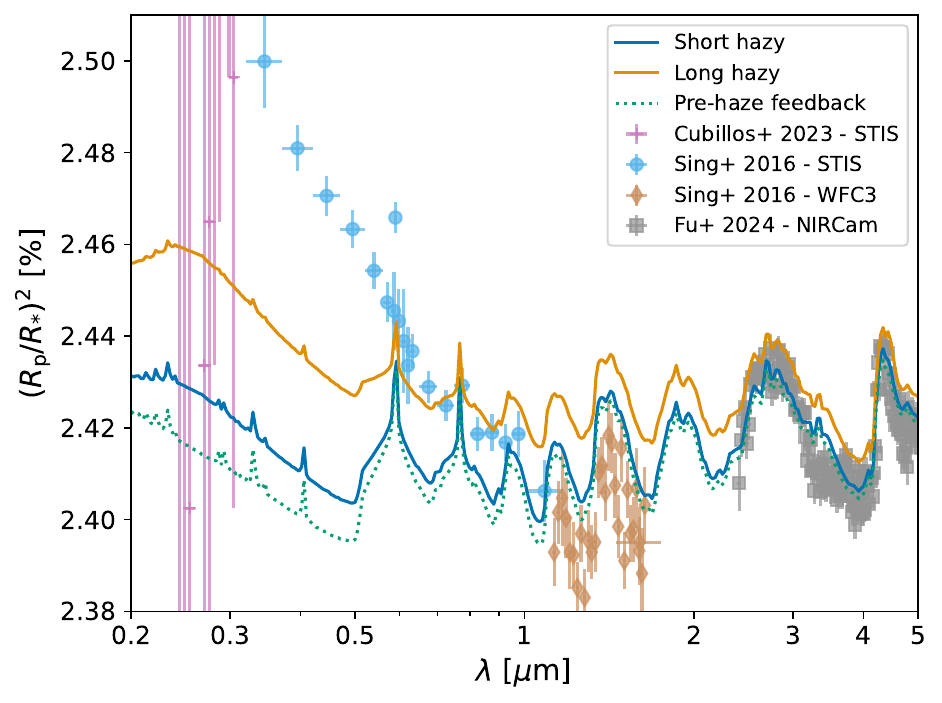}
    \includegraphics[width=0.49\linewidth]{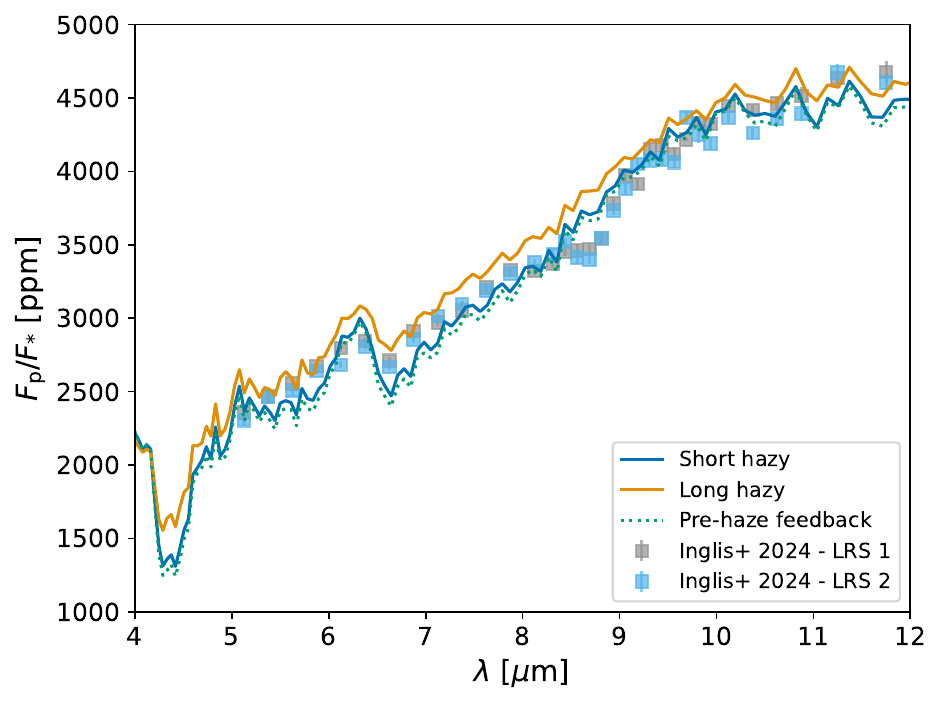}
    \caption{Transmission spectra (left), and dayside emission spectra (right) of our GCM simulations, for the short (blue solid line), and long (orange solid line) haze formation timescales.
    Green dotted lines show the spectrum before the radiative-feedback from haze was introduced.
    We include the available transmission spectra for HD 189733b from \citet{Cubillos2023}, \citet{Sing2016}, and \citet{Fu2024}, as well as the MIRI LRS dayside emission data from \citet{Inglis2024} for illustration.}
    \label{fig:trans_em}
\end{figure*}

\begin{figure*}
    \centering
    \includegraphics[width=0.49\linewidth]{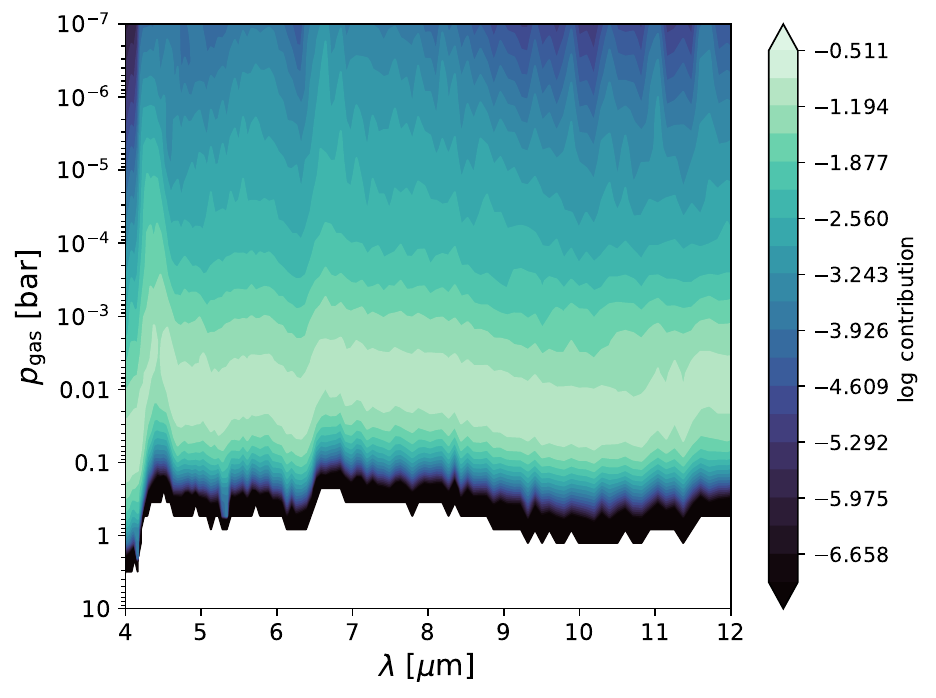}
    \includegraphics[width=0.49\linewidth]{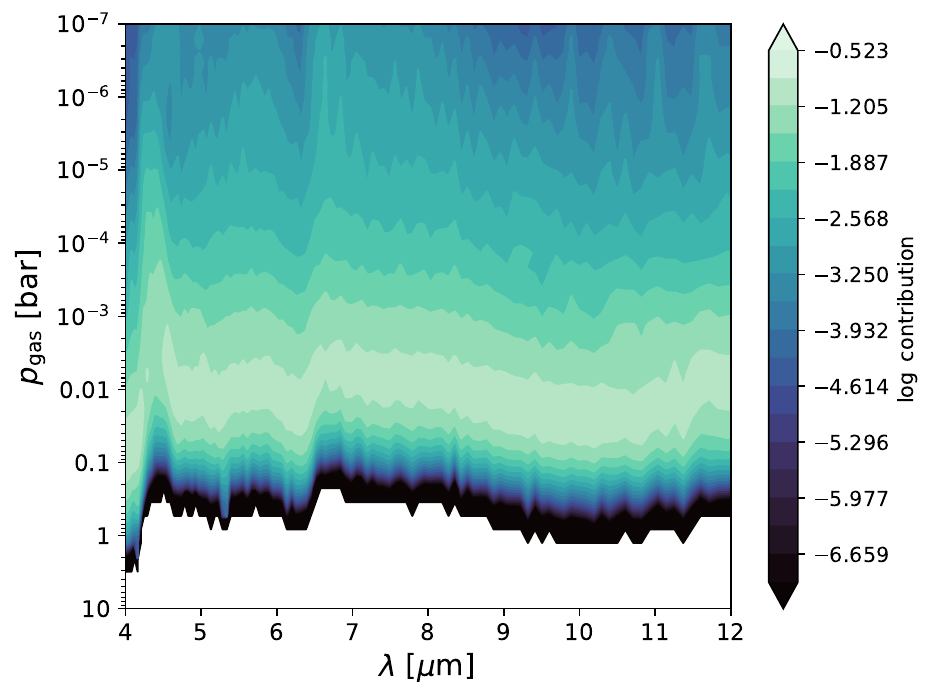}
    \caption{Dayside averaged 3D contribution functions of the GCM simulations for short (left) and long (right) haze formation timescale simulations.
    The regions contributing to the emission spectra are similar between each simulation.}
    \label{fig:con_fun}
\end{figure*}

In this Section, we post-process the results of the GCM simulations using the 3D radiative-transfer model {\tt gCMCRT} \citep{Lee2022}.
We produce synthetic transmission spectra and dayside emission spectra to explore the impact of the 3D haze distributions.
We assume a well peaked, lognormal, distribution with geometric standard deviation of $\sigma_{\rm g}$ = 1.05, taking the mean haze particle size (Eq. \ref{eq:rv}) in each cell as the median value for the distribution. 
This helps smooth the Mie theory calculations to avoid strong resonance bump features that can occur when assuming a single particle size.

Figure \ref{fig:trans_em} presents the transmission spectra of our GCM simulation output. 
In transmission, both our haze timescales simulation results add appreciable opacity to the atmosphere, resulting in muted spectral features across the infrared wavelength regime.
The long timescale results produce a more significant source of haze opacity, greatly flattening the spectra.
This is expected from our GCM results, where we found that larger thermal feedback effects were present for the long timescale simulation compared to the short timescale, suggesting a large overall haze opacity component.
This is probably due to the generally larger particle sizes in the long timescale simulation present across the globe.
However, similar to \citet{Steinrueck2023} and \citet{Mak2025}, we find that soot particles, without tuning the vertical mixing profiles, cannot fit the observed strong Rayleigh-like slope at optical wavelengths.

Figure \ref{fig:trans_em} also shows the dayside emission spectra at the mid-infrared regime, compared to the \citet{Inglis2024} JWST MIRI LRS measurements. 
We find that the haze particles generally increase the planetary flux due to the increased temperatures on the dayside of the atmosphere from thermal feedback of the haze opacity (Fig. \ref{fig:gcms}).
The short timescale simulations show only a slight increase in the dayside emission, showing that the temperature structures are generally similar with and without haze feedback.
However, the long timescale simulations show a stronger thermal response, raising the emission spectra by 100--200 ppm across the MIRI LRS wavelength regime.
This feedback places the spectrum generally too high compared to the MIRI LRS data from \citet{Inglis2024}.

Figure \ref{fig:con_fun} shows the 3D dayside averaged fractional contribution function for both the short and long timescale simulations for the MIRI LRS wavelength range.
Both simulations show similar contribution function contours, with most flux emanating between 0.1 and 0.01\,bar.
This region is where the maximum haze particle sizes are found (Figs. \ref{fig:vert_short} and \ref{fig:vert_long}), so that the dayside emission probes the same pressure range over which the haze opacity and its thermal feedback are strongest. 
This explains why the long timescale simulation, with its larger particles and stronger upper-atmosphere heating (Fig. \ref{fig:gcms}), produces the elevated dayside flux seen in Fig. \ref{fig:trans_em}, while the contribution function itself remains largely unchanged between the two cases: the haze alters the temperature at these levels rather than shifting the levels from which the flux emerges.

\section{Discussion}
\label{sec:disc}

In this study, we have parameterised the production rate of precursor molecules and a timescale dependent activation and haze formation scheme.
Several laboratory studies have now produced haze formation rates and compositions for sets of exoplanetary conditions \citep[e.g.][]{Horst2018, He2020, Moran2020, Thompson_2026}. 
Future modelling efforts can attempt to model atmospheres that contain similar compositions, but with the haze formation timescales estimated from the available experimental data.
This may be more feasible for sub-Neptune modelling where haze formation experiments in sub-Neptune-like atmospheric environments have already been performed in detail \citep[e.g.][]{Horst2018}.
This would be a good test of the modelling approach and provide a useful synergy between laboratory studies and theory.

In this study, we have assumed constant, highly parameterised activation and decay timescales for a generic precursor set of molecules.
Deriving temperature and pressure dependent rates for a set of precursors using photochemical models \citep[e.g.][]{Tsai2021}, possibly using simplifying methods such as chemical relaxation timescales \citep[e.g.][]{Tsai2018} may be warranted in the future.
These efforts would produce a more chemically consistent haze formation timescale for 3D simulations.
Most naturally, chemically consistent 3D kinetic-chemistry models such as {\tt mini-chem} \citep{Lee2023} or the \citet{Drummond2020,Zamyatina2023} scheme, as well as current 2D photochemical modelling efforts \citep[e.g.][]{Tsai2023} could be coupled to {\tt mini-haze} to investigate the effect of 2D/3D precursor molecule distributions on haze particle formation rates.
However, for the 3D models, photochemical processes would have to be included before a more fully consistent 3D approach would be feasible.

\citet{Lavvas2017} applied a haze evolution model originally designed for Titan studies \citep{Lavvas2010} to the atmosphere of HD 209458b and HD 189733b.
This uses a 1D bin evolution model to evolve the haze particle size distribution, including collisional growth processes and a similar parameterisation for the height dependent haze formation rate to that used here.
Although it is not straightforward to compare 1D simulations with 3D simulations, our results are in good agreement with the structures and particle sizes found by \citet{Lavvas2017} at similar haze particle formation rates.
For example, \citet{Lavvas2017} HD 189733b simulation produces haze particle sizes up to $\sim$ 20--30\,nm for their 10$^{-12}$\,g\,cm$^{-2}$\,s$^{-1}$, 0.1 $K_{\rm zz}$ profile case, which is in line with the GCM results which have an overall similar haze particle formation rate.
We suggest that since the particle sizes remain relatively small in this particular haze formation scenario, the monodisperse approximation of the moment method holds well and strong polydisperse components are not important in altering the growth rates significantly.
The results from the bin model \citet{Gao2023} suggest some polydisperse behaviour can occur but the bulk of the haze mass is retained in the small particle sizes of $<$ 0.1\,$\mu$m.
Overall, the similarity between the vertical haze structures and particle sizes in the complex \citet{Lavvas2017} bin model and our GCM moment approach, suggests that our moment method is accurately capturing the salient haze collisional growth processes in the atmosphere.

\section{Conclusions}
\label{sec:conc}

In this study, we presented `{\tt mini-haze}', a flexible two-moment photochemical haze formation scheme designed for ease of use in time-dependent exoplanet atmosphere simulations.
{\tt mini-haze} evolves the haze particle number density and particle size properties self-consistently, using a coagulation-coalescence approach, going beyond the previous single-sized haze particle studies used by \citet{Steinrueck2021} and \citet{Steinrueck2023}.
{\tt mini-haze} is available online on GitHub\footnote{\url{https://github.com/ELeeAstro/mini_haze}}. 

Inspired by recent 2D kinetic chemistry \citep{Tsai2023} and cloud particle condensation models \citep{Powell2024}, we propose a simple activation and decay chemical timescale method to emulate the time evolution and efficiency of precursor molecule conversion into solid haze particles.
Through choosing the balance between these timescales, the appropriate conversion efficiency between precursor molecules and haze particles can be naturally included in the model.
Our current model uses a simple parameterised pressure dependent haze production rate similar to previous modelling efforts.
Our scheme is flexible for future efforts that may self-consistently calculate a production rate based on the 3D time-dependent chemical structure.

In an initial test, we apply our model to simulate a hazy hot Jupiter HD 189733b atmosphere scenario using {\tt mini-haze} coupled to the {\tt Exo-FMS} GCM.
We examine differences between assuming an instantaneous haze formation timescale compared to a delayed timescale, approximately half the advective timescale, to allow a component of the haze to form on the nightside hemisphere of the planet.
Our chosen precursor formation rate results in maximum haze particle sizes of $\sim$ 2--30\,nm, with the 3D distribution of hazes closely following the dominant horizontal dynamical structures and vertical velocity profiles of the atmosphere, as seen in previous 3D hot Jupiter haze particle studies \citep{Steinrueck2023}.

Our transmission spectra results show muting of spectral features from the presence of haze particles, with the long formation timescale haze producing appreciably more opacity than the short formation timescale results.
We find the emission spectrum is not affected significantly from the presence of haze particle opacity itself, but the stronger thermal feedback present in the long timescale simulation compared to the short timescale simulation leads to an overall greater planetary flux.

Our current model conforms well with previous 1D studies that used a complex bin haze formation model on the same planet \citep{Lavvas2017}, reproducing similar particle number densities and particle sizes at comparable haze formation rates.
The coupled {\tt mini-haze} and GCM models reproduce well the global haze distribution of studies that used parameterised haze particle sizes \citep{Steinrueck2021, Steinrueck2023}, with the 3D distribution of haze particles following the similar vertical velocity structures, vertical mixing ratios, thermal feedback effects and overall effect on transmission and emission spectra.
While our results are complementary and concordant with \citet{Steinrueck2021}, \citet{Steinrueck2023} and \citet{Mak2025}, the key advance of {\tt mini-haze} is that the particle sizes are evolved self-consistently through collisional growth rather than fixed a priori, and that the haze precursor production and conversion efficiency are set by a physically motivated timescale scheme rather than a prescribed solid formation rate.

Overall, {\tt mini-haze} offers a general, flexible and efficient way of including a haze formation component in time-dependent models of exoplanet atmospheres.
{\tt mini-haze} uses a two moment method, as well as a simple precursor material to haze particle timescale scheme, to model the production of haze monomers and their subsequent collisional growth across the atmosphere.
{\tt mini-haze} is highly complementary to current and future photochemical models that aim to add a haze formation component to their kinetic chemistry models.
Our current methodology including haze formation and evolution, kinetic chemistry and thermal-feedback effects represents a step towards full self-consistency in understanding chemical and haze interactions in exoplanet atmospheres.
In a forthcoming paper (Paper II), we will extend the {\tt mini-haze} framework to cooler sub-Neptune atmospheres and explore how different haze compositions and formation efficiencies shape their 3D spatial distributions and observable spectra.

\begin{acknowledgements}
We thank P. Lavvas for making available soot particle optical constants.
E.K.H. Lee is supported by the CSH Bernoulli Fellowship.
M.E. Steinrueck is supported by a 51 Pegasi b fellowship from the Heising-Simons Foundation.
K. Ohno acknowledges support from the JSPS KAKENHI grant numbers JP23K19072 and JP21H01141.
X. Zhang acknowledges support from the NSF grant (AST2307463), NASA Exoplanet Research grant (80NSSC22K0236), and the NASA Interdisciplinary Consortia for Astrobiology Research grant (80NSSC21K0597). 
This work benefited from the 2024 Exoplanet Summer Program in the Other Worlds Laboratory (OWL) at the University of California, Santa Cruz, a program funded by the Heising-Simons Foundation and NASA. 
\end{acknowledgements}

\bibliographystyle{aa} 
\bibliography{bib.bib}

\appendix

\section{Longitude-latitude maps}

\begin{figure*}
    \centering
    \includegraphics[width=0.45\linewidth]{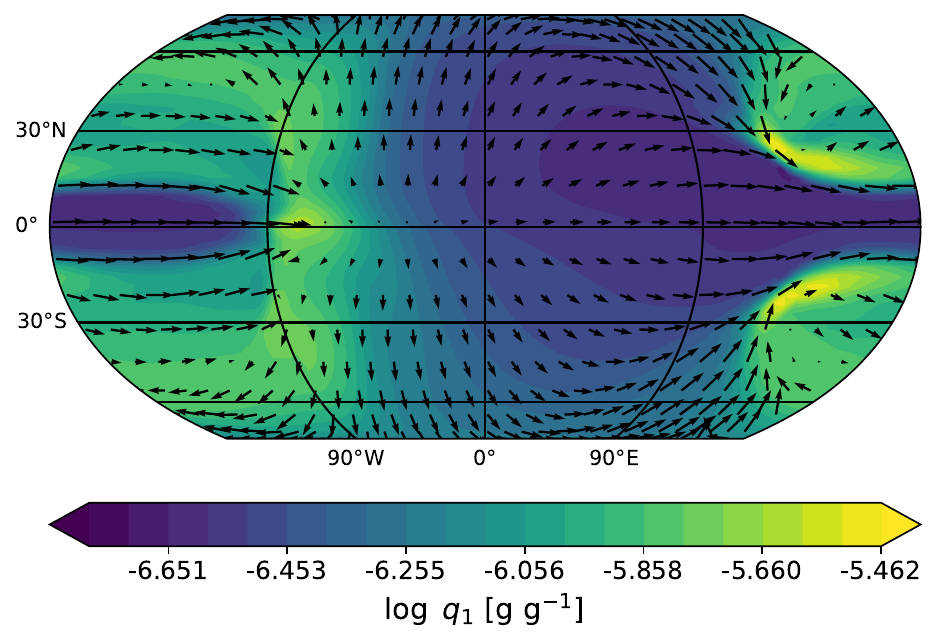}
    \includegraphics[width=0.45\linewidth]{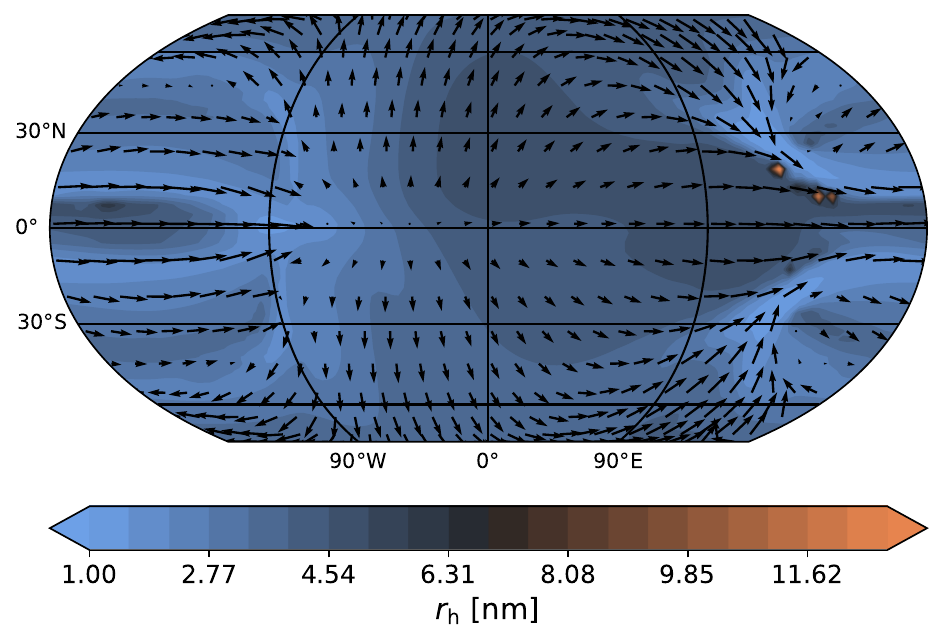}
    \includegraphics[width=0.45\linewidth]{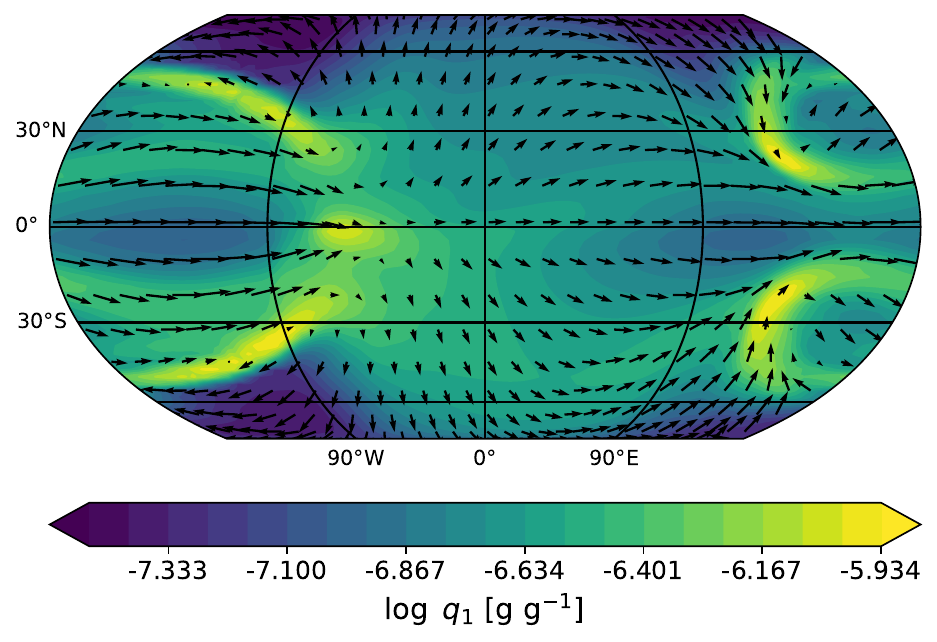}
    \includegraphics[width=0.45\linewidth]{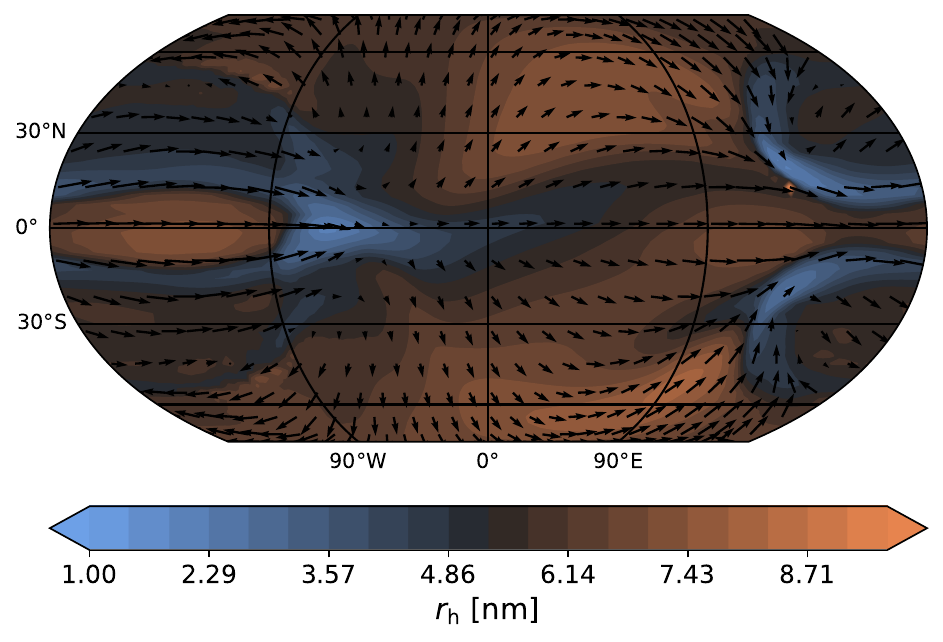}
    \includegraphics[width=0.45\linewidth]{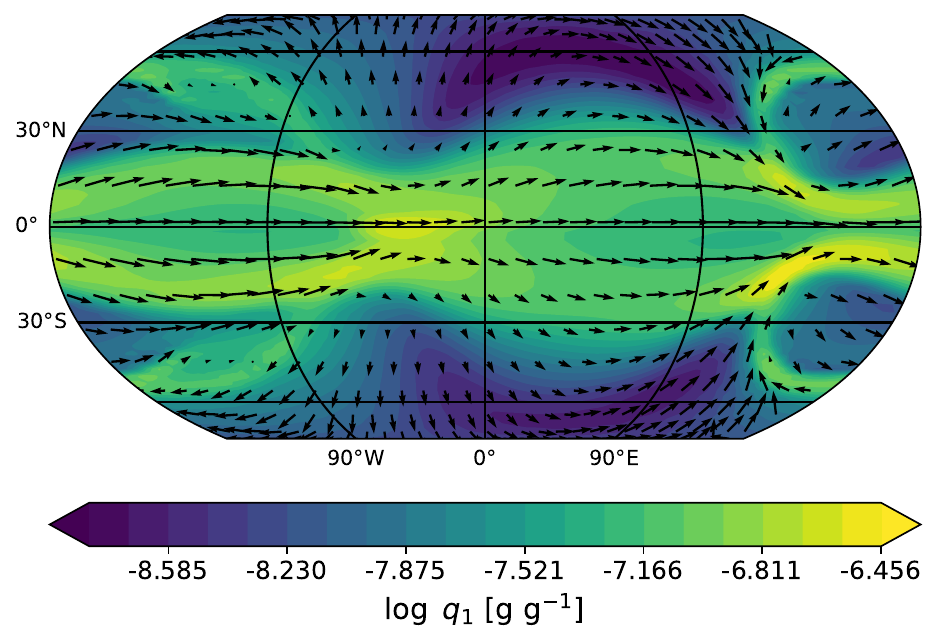}
    \includegraphics[width=0.45\linewidth]{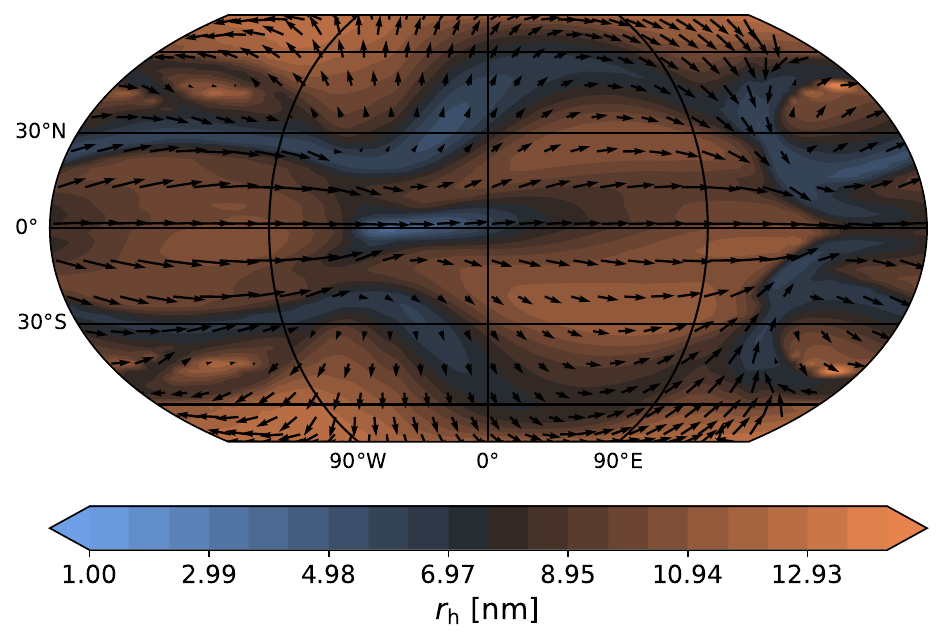}
    \includegraphics[width=0.45\linewidth]{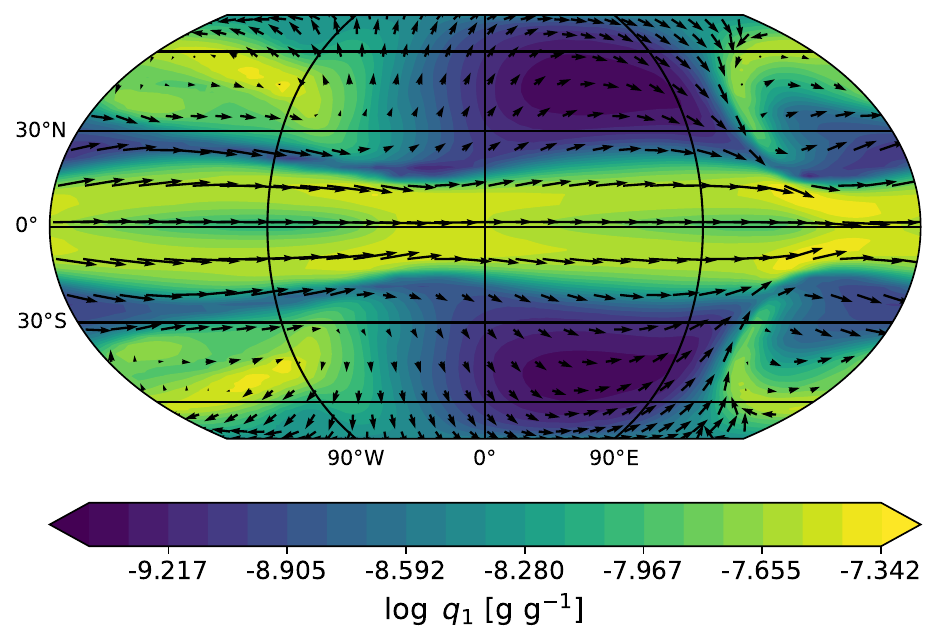}
    \includegraphics[width=0.45\linewidth]{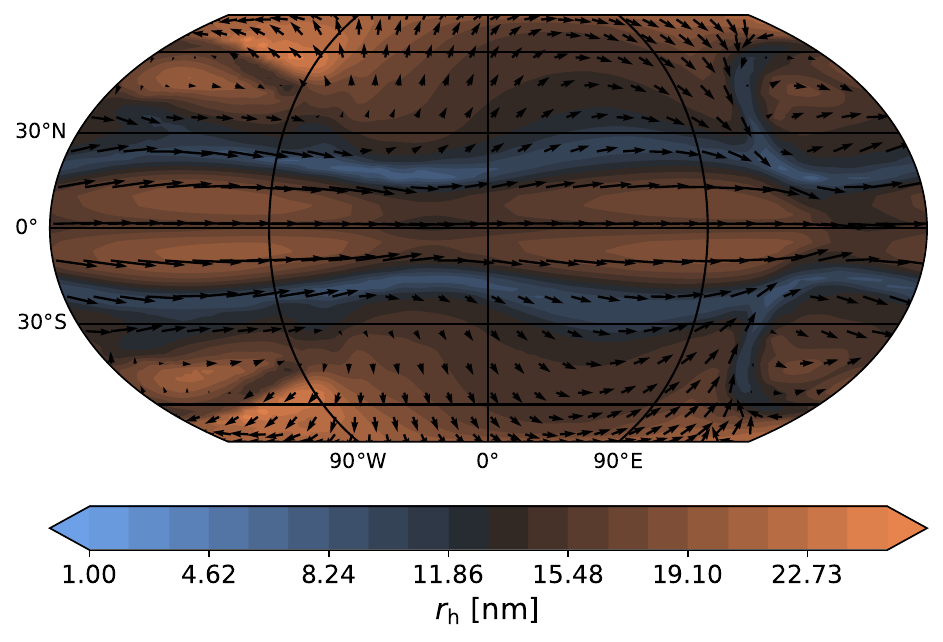}
    \caption{Haze mass mixing ratio, $q_{1}$ [g g$^{-1}$] (left column), and haze particle size , $r_{\rm h}$ [nm] (right column), for the short formation timescale simulation at 10$^{-5}$\,bar (top row), 10$^{-4}$\,bar (second row), 10$^{-3}$\,bar (third row) and 10$^{-2}$\,bar (bottom row) pressure levels.
    The sub-stellar point is located at (0$^{\circ}$, 0$^{\circ}$).}
    \label{fig:map_short}
\end{figure*}

\begin{figure*}
    \centering
    \includegraphics[width=0.45\linewidth]{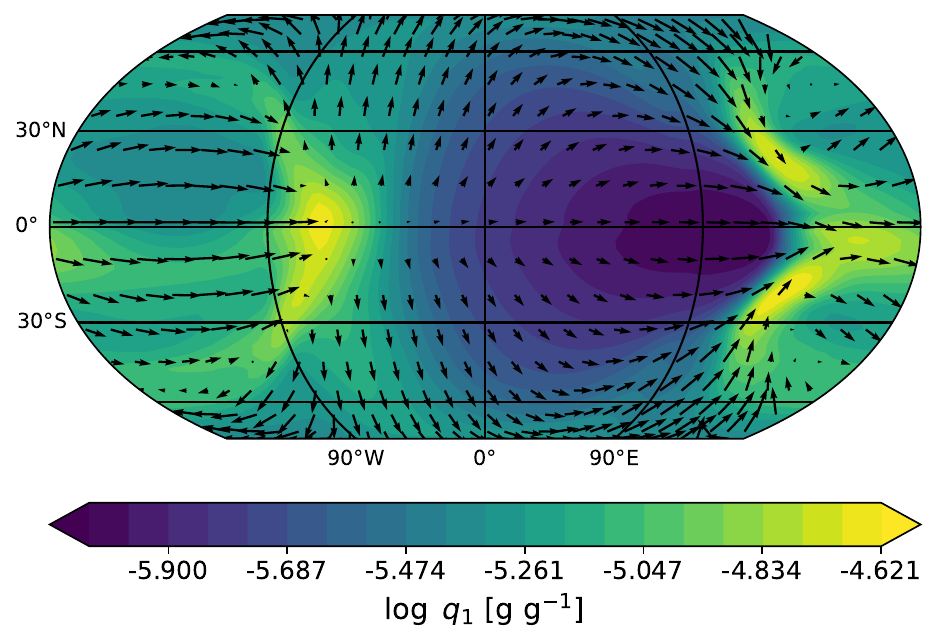}
    \includegraphics[width=0.45\linewidth]{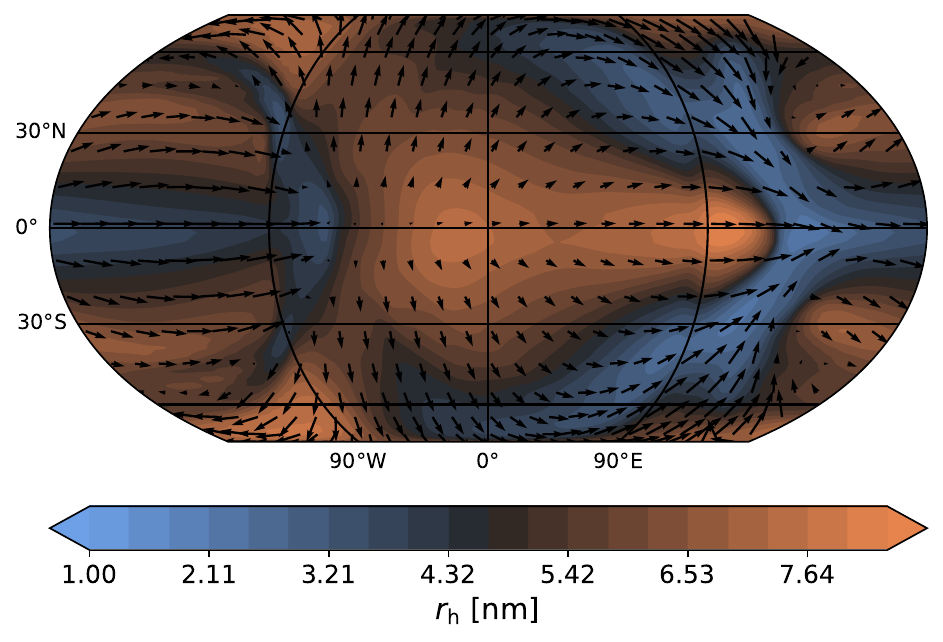}
    \includegraphics[width=0.45\linewidth]{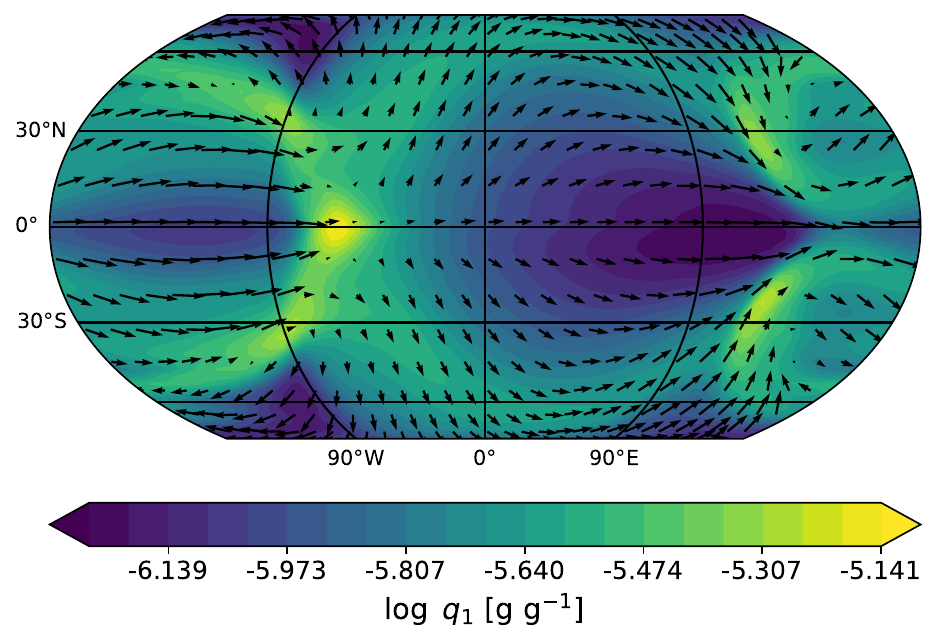}
    \includegraphics[width=0.45\linewidth]{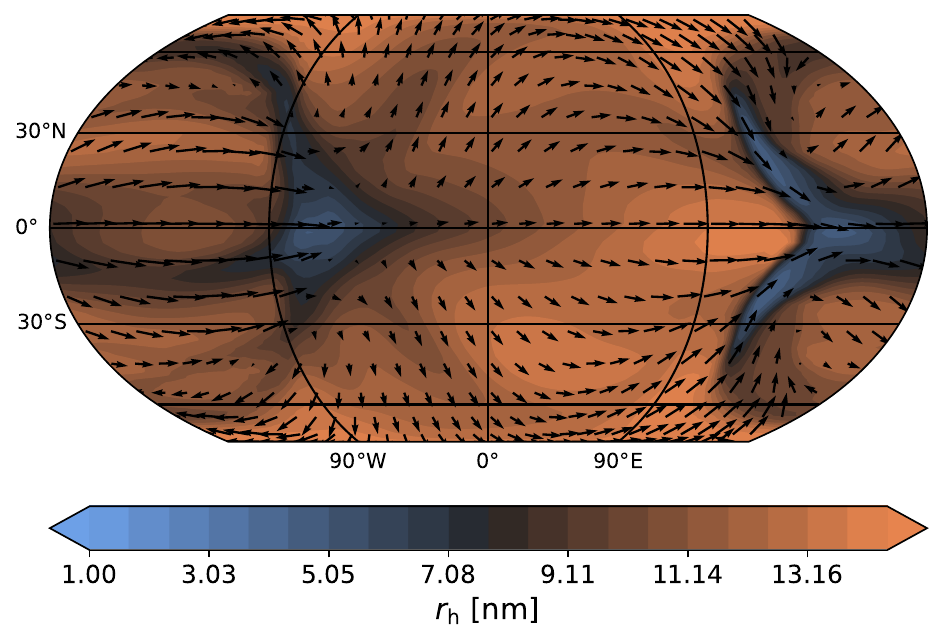}
    \includegraphics[width=0.45\linewidth]{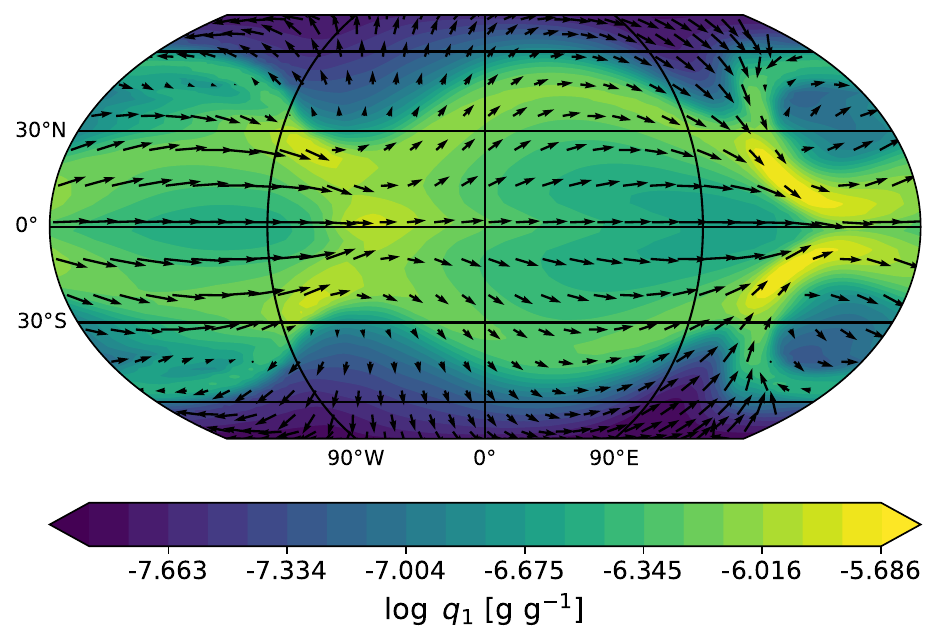}
    \includegraphics[width=0.45\linewidth]{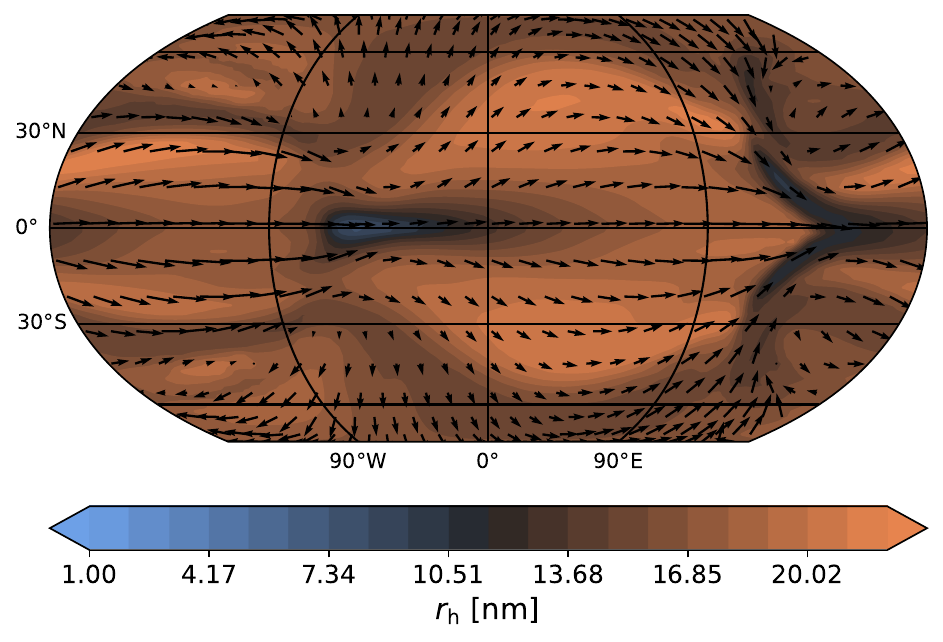}
    \includegraphics[width=0.45\linewidth]{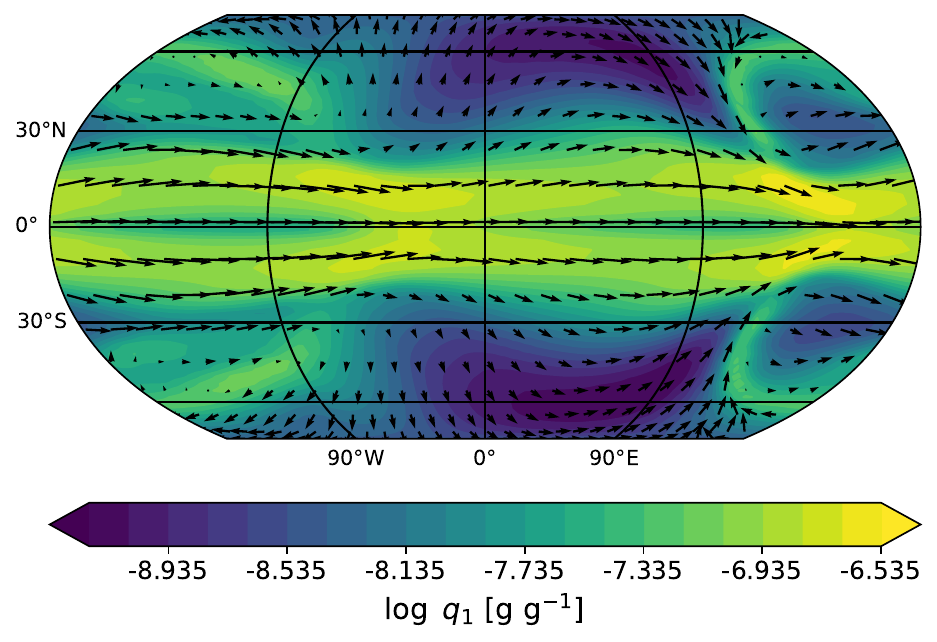}
    \includegraphics[width=0.45\linewidth]{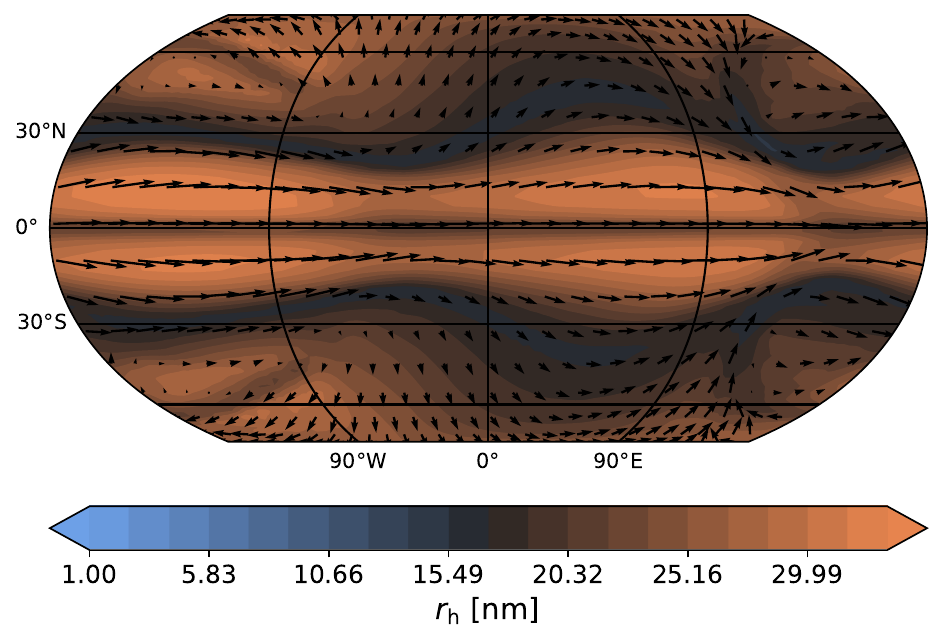}
    \caption{Same as Fig. \ref{fig:map_short} but for the long haze formation timescale.}
    \label{fig:map_long}
\end{figure*}

In this Appendix, we present the latitude-longitude maps of each simulation at various pressure levels for the short (Fig. \ref{fig:map_short}) and long (Fig. \ref{fig:map_long}) haze formation timescales.

\end{document}